\documentclass[twocolumn]{aastex63}

\usepackage{graphics}
\usepackage{float}
\usepackage{comment}
\usepackage{aas_macros}

\newcommand{\LA}{L$_{\rm A}$}

\newcommand\Mach{\mbox{Ma}}

\revised{\today}
\shorttitle{3D simulations in Si and O-rich Layers}
\shortauthors{Yoshida et al.}

\begin{document}

\title{Three-dimensional Hydrodynamics Simulations of Precollapse Shell Burning in the Si- and O-rich Layers }

\correspondingauthor{Takashi Yoshida}
\email{tyoshida@astron.s.u-tokyo.ac.jp}

\author[0000-0002-8967-7063]{Takashi Yoshida}
\affil{Department of Astronomy, Graduate School of Science, University of Tokyo, 7-3-1 Hongo, Bunkyo-ku, Tokyo 113-0033, Japan}

\author[0000-0003-0304-9283]{Tomoya Takiwaki}
\affiliation{Division of Science, National Astronomical Observatory of Japan, 2-21-1 Osawa, Mitaka, Tokyo 181-8588, Japan}

\author[0000-0003-2456-6183]{Kei Kotake}
\affiliation{Department of Applied Physics \& Research Institute of Stellar Explosive Phenomena, Fukuoka University, Fukuoka 814-0180, Japan}

\author[0000-0002-6705-6303]{Koh Takahashi}
\affiliation{Max Planck Institute for Gravitational Physics, D-14476 Potsdam, Germany}

\author[0000-0002-8734-2147]{Ko Nakamura}
\affiliation{Department of Applied Physics, Fukuoka University, Fukuoka 814-0180, Japan}

\author[0000-0001-8338-502X]{Hideyuki Umeda}
\affiliation{Department of Astronomy, Graduate School of Science, University of Tokyo, 7-3-1 Hongo, Bunkyo-ku, Tokyo 113-0033, Japan}

\begin{abstract}
We present 3D hydrodynamics simulations of shell burning in two progenitors
 with zero-age main-sequence masses of 22 and 27 $M_{\odot}$ for $\sim$65 and 200 s up to the onset of gravitational collapse, respectively. 
The 22 and 27 $M_{\odot}$ stars are selected from a suite of 1D progenitors.
The former and the latter have an extended Si- and  O-rich layer with a width of $\sim$10$^9$ cm and $\sim$5$\times 10^9$ cm, respectively.
Our 3D results show that turbulent mixing occurs in both of the progenitors with the angle-averaged turbulent Mach number exceeding $\sim$0.1 at the maximum.
We observe that an episodic burning of O and Ne, which takes place underneath the convection bases, enhances the turbulent mixing in the 22 and 27 $M_\odot$ models, respectively.
The distribution of nucleosynthetic yields is significantly different from that in 1D simulations, namely, in 3D more homogeneous and inhomogeneous in the radial and angular direction, respectively.
By performing a spectrum analysis, we investigate the growth of turbulence and its role of material mixing in the convective layers. 
We also present a scalar spherical harmonics mode analysis of the turbulent Mach number.
This analytical formula would be helpful for supernova modelers to implement the precollapse perturbations in core-collapse supernova simulations.
Based on the results, we discuss implications for the possible onset of the perturbation-aided neutrino-driven supernova explosion.
\end{abstract}


\section{Introduction} \label{sec:intro}

Multidimensional effects in the late burning stages of massive stars have received considerable attention for energizing neutrino-driven supernova (SN) explosions.
\citet{couch_ott13} first demonstrated that inhomogeneities seeded by convective shell burning assist the onset of a neutrino-driven explosion \citep[see also][]{rodorigo14,takahashi14,couch_ott15,bernhard15,Burrows18,nagakura19}. The
 primary reason of the perturbation-aided explosion is that the infalling perturbation   enhances turbulence behind the postshock matter, leading to the reduction of the critical neutrino luminosity for shock revival \citep[e.g.,][]{bernhard15,ernazar16}. 
In these studies, the nonspherical structures in the burning shells were treated in a parametric manner. 
This is because of the paucity of multidimensional stellar evolution models covering the life span of massive stars up to the iron core collapse (CC). 
Currently 1D stellar evolution calculations are the only way to do this \citep{woosley02,Woosley07,Sukhbold18}, where multidimensional effects are phenomenologically treated by the parameters of the mixing length theory (MLT) \citep[e.g.,][]{kippenhahn12}. 
 
Since the 1990s, 2D and 3D stellar evolution simulations focusing on the late burning stages have been extensively conducted \citep{arnett94,bazan94,bazan98,asida00,kuhlen03,meakin06,meakin07,arnett11,chatz14,chatz16,jones17}. 
The multidimensional hydrodynamics stellar evolution calculations have been done over several turnover timescales of convection (limited by the computational resources) in selected burning shells (e.g., \citet{meakin07}; \citet{viallet13,simon15,cristini17}; \citet{cristini19,and20} for different burning shells; see \citet{arnett16} for a review).
  
Recently, several important attempts to evolve convective shells in 3D prior to the onset of collapse have been reported by \citet{Couch15} for silicon (Si) shell burning in a 15 $M_{\odot}$ star and by \citet{bernhard16_prog} for oxygen (O) shell burning in an 18 $M_{\odot}$ star.
\citet{Couch15} obtained earlier onset of a neutrino-driven explosion for the 3D progenitor model compared to the corresponding 1D model \citep[see also][]{bernhard17}.
More recently, a 3D simulation of a shell merger
 of convective O and Ne layers has been reported in \citet{Yadav19} for an $18.88 M_{\odot}$ star  \citep[see also][]{mocak18}.
  \citet{Yadav19} were the first to point out that the shell merger could explain asymmetric features in SN remnants such as the Si--Mg rich "bullet"-like features in Cassiopeia A (e.g., \citet{gref17}, see also \citet{annop17,utrobin19,ono20} for collective references therein).
These studies present evidence that 3D modeling of the convective layers is necessary not only for verifying a universality of the perturbation-aided explosion but also for probing into the multidimensional stellar evolution hydrodynamics features (such as the shell merger) by comparing with the observed nucleosynthetic yields.

Joining in these efforts, we investigated in our previous study \citep{Yoshida19} how the asphericities could grow, particularly driven by
the convective oxygen shell burning in the O- and Si-rich layer.  
First, we performed a series of 1D stellar evolution calculation with zero-age main-sequence (ZAMS) masses between 9 and 40 $M_{\odot}$. 
Based on the 1D results, we selected 11 1D progenitor models that have extended and enriched O and Si
layers, because they were expected to result in vigorous convection \citep[e.g.,][]{bernhard16_prog}.
Mapping the 1D progenitors to our multidimensional hydrodynamics
code \citep{Nakamura16,Takiwaki16}, we followed the 2D  evolution over a time of $\sim$100 s before the onset of collapse. 
Among the 11 2D models, we chose one progenitor of a $25M_{\sun}$ star that showed highest convective activity. 
We then followed the 3D evolution of the $25 M_{\odot}$ model focusing on the convection activity in the silicon- and oxygen-rich (Si/O-rich) layer up to the CC. We found that the 3D model develops large-scale ($\ell = 2$) convection similar to the 2D model; however, the turbulent velocity was lower in 3D than in 2D. 
 
As a sequel to our previous paper, we present results of two more 3D stellar evolution calculations in this work:
one is a 22 $M_{\odot}$ star with an extended Si/O-rich layer \citep[similar to that of the 25 $M_{\odot}$ model;][]{Yoshida19}, and another is a 27 $M_{\odot}$ star with a more extended oxygen- and silicon-rich (O/Si) layer than the 22 and 25 $M_{\odot}$ models. 
The reason of our choice of the 27 $M_{\odot}$ star is that the high Mach number region extends most farther out among our 2D models \citep[see the bottom right panel of Figure 5 in][]{Yoshida19}.
As pointed out by \citet{Yadav19}, such widely mixing layers could be of potential interest in the (yet-uncertain) nucleosynthesis context, which should be studied in 3D simulation. 
By computing the two new progenitors in 3D, we make a comparison of convective motions in the burning shells, turbulent Mach number, and typical scales of turbulent eddies. 
For this, we perform a spectrum analysis of the turbulent velocity. 
We also present results of a scalar spherical harmonics (SSH) expansion of the radial Mach number. To increase the sample number of 3D progenitor models, the 25 $M_{\odot}$ model result is analyzed as well. We hope that SN modelers can play with the precollapse inhomogeneities in CC SN simulations by utilizing the analytical formula provided in this work.

The paper is organized as follows.
Section \ref{sec2} starts with a brief description of our initial models of the 22 and 27 $M_{\odot}$ stars, 
as well as the numerical methods of our 3D stellar evolution calculation.
In Section \ref{sec3}, we present results of the two 3D stellar evolution models following about 65 s for the 22 $M_\odot$ star and 200 s for the 27 $M_\odot$ star up to the onset of CC.
In Section \ref{sec4}, we summarize with a discussion of the possible implications.

\begin{figure*}
\plottwo{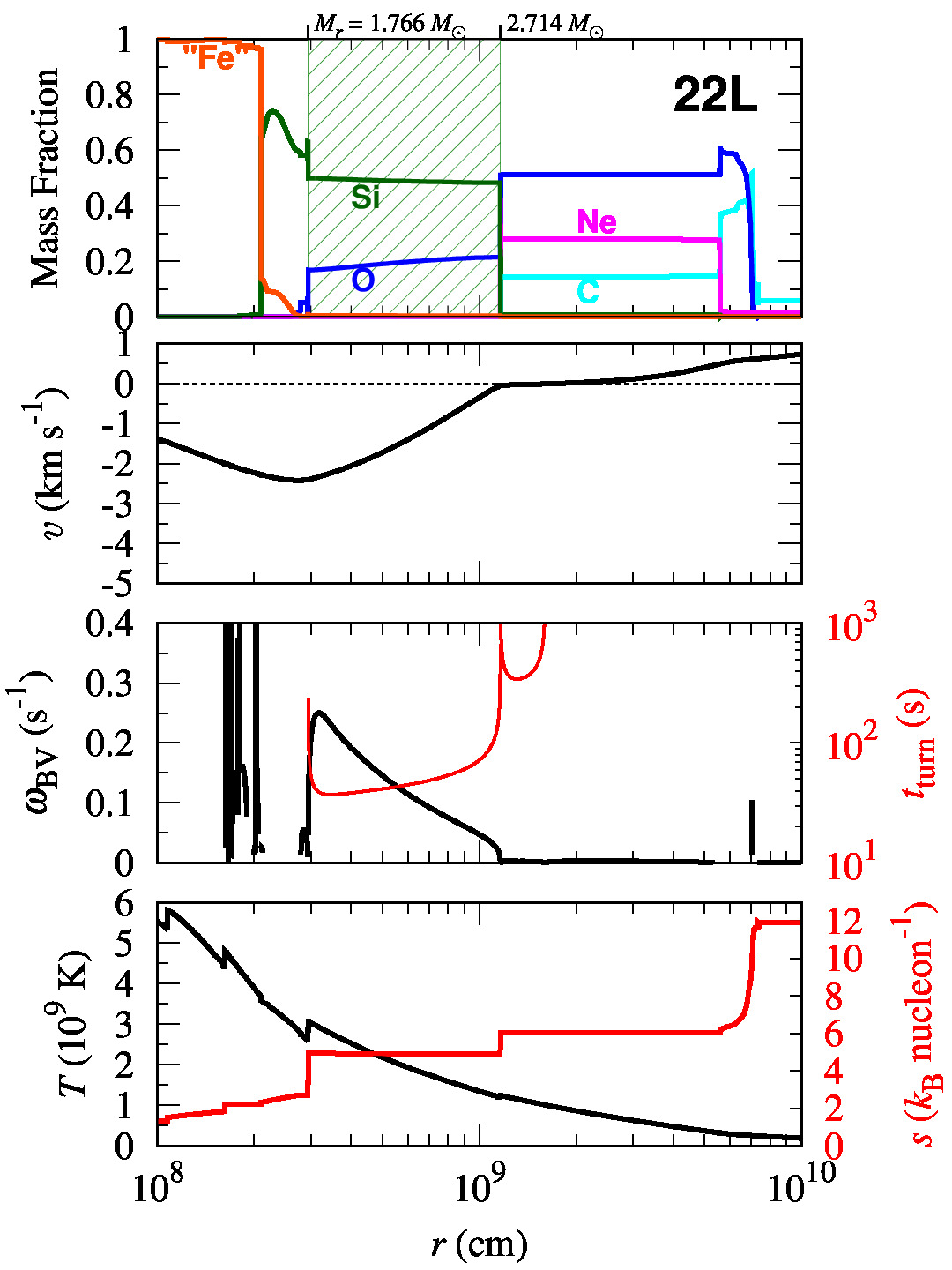}{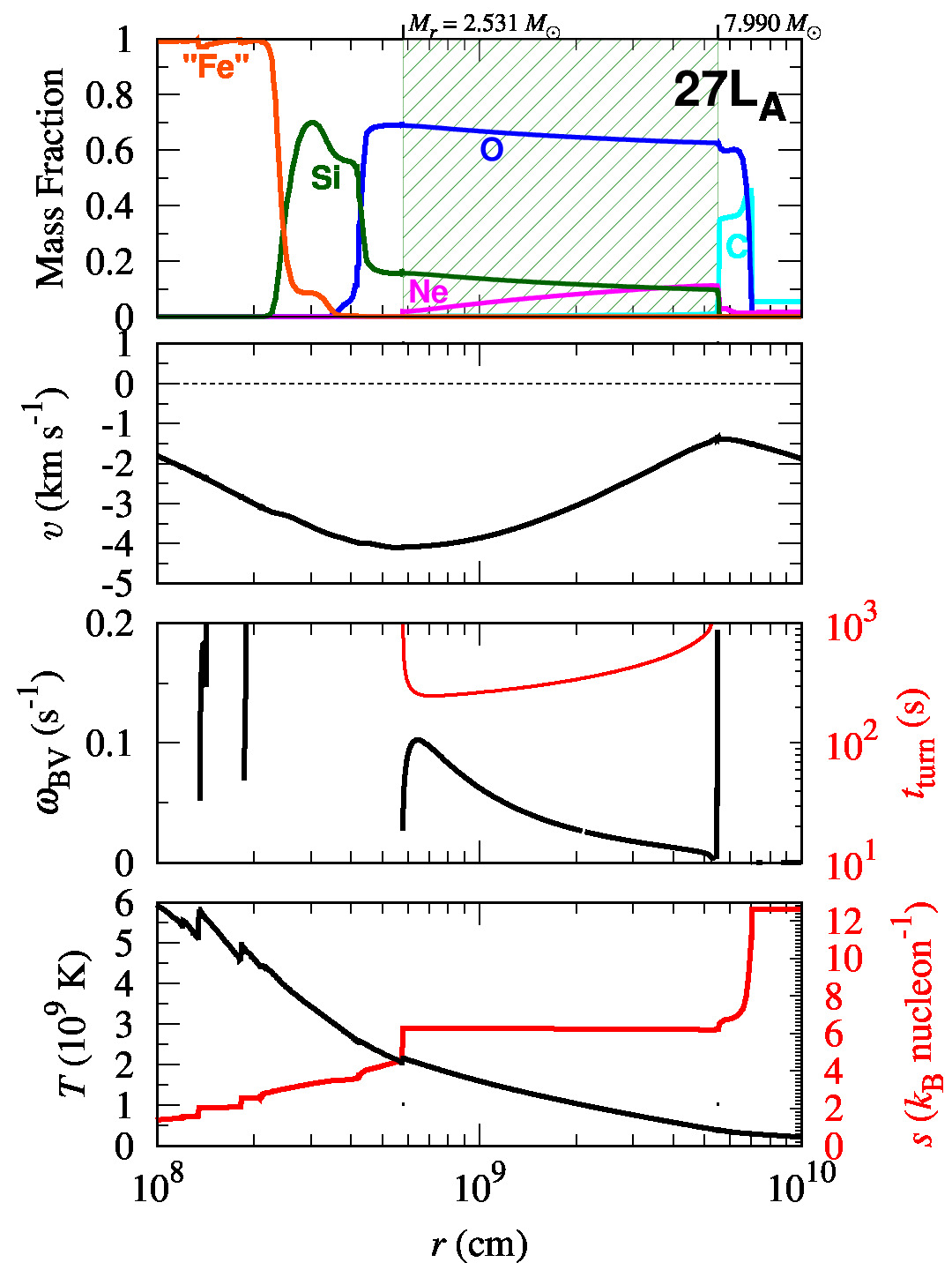}
\caption{
Initial structure for the 3D simulations of models 22L (left panels) and 27\LA (right panels).
Top row: initial mass fraction profiles.
Green hatched regions are Si/O-rich convective layers in spherically symmetrical stellar evolution calculations.
Second row: The radial velocity profiles.
Third row: Brunt-V\"{a}is\"{a}l\"{a} frequency $\omega_{\rm BV}$ (s$^{-1}$) and the turnover timescale of convection $t_{\rm turn}$ (s).
Bottom row: the radial profiles of temperature $T$ in units of $10^{9}$ K and entropy $s$ in units of $k_{\rm B} $ (the Boltzmann constant) per nucleon.
\label{fig:massfraction}}
\end{figure*}

\section{Initial Model and Numerical Method} \label{sec2}

In Section \ref{subsec:initial}, we start to describe the initial models. 
Then, we briefly summarize the numerical schemes of our 3D hydrodynamics simulations in Section \ref{subsec:hydro}.

\subsection{Initial Models} \label{subsec:initial}

As already mentioned above, we employ two 1D progenitors from \citet{Yoshida19}: one is a 22 $M_{\odot}$ star with an extended Si/O-rich layer (hereafter model 22L), and the other is a 27 $M_{\odot}$ star with a more extended O--Si layer (hereafter model 27\LA) than model 22L. 
Note that "L" in the model name represents the choice of the overshooting parameter, which is calibrated to explain early B-type stars in the Large Magellanic Cloud (LMC) \citep{Brott11}, and the subscript ``A" denotes the inclusion of a convective overshoot during the advanced stages\footnote[6]{The overshoot is taken as a diffusive manner. The diffusion coefficient of the overshoot is taken as $D_{\rm ov} = D_{\rm cv,0} \exp (-2\Delta r/f_{\rm ov} H_{P0})$, where $D_{\rm cv,0}$, $\Delta r$, $f_{\rm ov} = 0.002$, and $H_{P0}$ are the diffusion coefficient at the edge of the convective region, the distance from the edge, the overshoot parameter in advanced stage, and the pressure scale height at the edge, respectively.}.
Both models are nonrotating
(see \citet{Yoshida19} and \citet{Takahashi16,Takahashi18} for more details about the 1D stellar evolution code ``HOSHI" ).

Figure \ref{fig:massfraction} shows the initial structure for the 3D simulations of model 22L (left panel) and model 27\LA (right panel).
These initial structures correspond to the 1D structures at $\sim$130 s before the central temperature is $10^{10}$ K.
In four rows from the top to the bottom, we show the mass fraction profiles, the radial velocity profiles, 
the profiles of Brunt-V\"{a}is\"{a}l\"{a} frequency and turnover time of the convection, and the temperature and entropy profiles.
Brunt-V\"{a}is\"{a}l\"{a} frequency in convective layers is defined as
\begin{equation}
    \omega_{\rm BV} = \sqrt{- \frac{GM_r}{r^2} \frac{\delta}{H_P} (\nabla_{\rm ad} - \nabla + \frac{\varphi}{\delta} \nabla_{\mu} )},
\end{equation}
where $G$ is the gravitational constant, $M_r$ is the mass coordinate, $r$ is radius, 
$\delta \equiv -(\partial  \ln \rho/\partial \ln T)_{P,\mu}$, $H_P$ is pressure scale height, 
$\nabla_{\rm ad} \equiv (\partial \ln T/\partial \ln P)_{\rm ad}$ is the adiabatic temperature gradient, 
$\nabla \equiv \partial \ln T/\partial \ln P$ is the temperature gradient, 
$\varphi \equiv (\partial \ln \rho/\partial \ln \mu)_{P,T}$, and $\nabla_{\mu} \equiv \partial \ln \mu/\partial \ln P$.
Note in this definition that $\omega^2_{\rm BV} > 0$ corresponds to convective instability.
The turnover time of convection in a convective layer is calculated using the equation $t_{\rm turn} = r_{\rm cv}/v_{\rm cv}$, where $r_{\rm cv}$ is the width of the convection layer and $v_{\rm cv}$ is the convection velocity calculated using MLT.

The left panel shows the initial structure of model 22L.
The top row indicates the mass fraction profiles.
The green hatched region shows the location of the Si/O layer  ($ 3 \times 10^{8}$ cm $\leq r \leq 1.2 \times10^{9}$ cm and  1.77 $M_\odot \leq M_r \leq 2.71 M_\odot$).
The width of the Si/O layer is $9 \times 10^{8}$ cm.
This layer is convective owing to O shell burning. 
In fact, the temperature of the base of the layer (at $r \sim 3 \times 10^{8}$ cm, the left end of the hatched region) is about $3 \times 10^{9}$ K (see the black line in the bottom left panel), which exceeds the ignition temperature of O burning. 
The entropy profile (red line in the bottom left panel) is nearly flat in the Si/O layer (shaded region), which is a result of convective energy transport treated in the 1D stellar evolution calculation. 
Note that the regions inside and outside the shaded region correspond to the Fe/Si core and O/Ne layer, respectively.  
Both of the regions are convectively inert in the 3D simulation. 
The former is due to a positive entropy gradient in the core; the latter is due to too short computational timescale for our 3D simulation to see the convective activity for the outer region (i.e., longer convective timescale; see the third row panel).

Note that the composition structure of model 22L is similar to model 25M, of which 2D and 3D hydrodynamics simulation was investigated in \citet{Yoshida19}. 
In the previous study, the 2D dynamical simulation has also been done for the model 22L, which gives a similar result to the 2D simulation of the model 25M.
The maximum turbulent Mach number reaches 0.108, and the spectrum analysis showed a peak of the radial turbulent velocity at a low mode of $\ell = 2$. 
For model 22L, we start the 3D calculation mapping from the 1D data when the temperature and density at the center are $6.40 \times 10^{9}$ K and $1.32 \times 10^{9}$ g cm$^{-3}$, respectively. We follow the 3D evolution for 65.5 s until the onset of CC.

The right panel is for model 27L$_{\rm A}$.
It has an extended O/Si-rich layer at $5\times 10^{8}$ cm $\leq r \leq 60 \times 10^{8}$ cm 
(2.53 $M_\odot \leq M_r \leq 7.99 M_\odot$).
The inner neon-depleted region ($r < 5.8 \times 10^8$ cm) is formed by oxygen shell burning after oxygen core burning.
The outer region ($r > 5.8 \times 10^{8}$ cm) with a small amount of neon is the former O/Ne layer. Neon in this region 
($5.8 \times 10^{8}$ cm $\le r \le 60 \times 10^{8}$ cm) is burned through Ne shell burning after the Si core burning phase but still remains after the Fe core formation.
We call this layer the O/Si/Ne layer, which is indicated by the green hatched region. 
The O/Si/Ne layer is convective and has constant entropy profile (red line). The shell convection is powered by Ne shell burning, as the base temperature exceeds the ignition temperature of Ne burning of $2 \times 10^9$ K (black line).

The 2D model of 27L$_{\rm A}$ showed the maximum turbulent Mach number 0.179 with a peak mode of the radial Mach number $\ell = 2$ in \citet{Yoshida19}.  
The convective region extends to the whole O/Si/Ne layer. 
Note that the region outside and inside of this layer is convectively inert in the 3D simulation owing to the same reason as for model 22L as mentioned above. For this model, we start the 3D calculation by mapping from the 1D data when the temperature and density at the center are $6.87 \times 10^{9}$ K and $9.44 \times 10^{8}$ g cm$^{-3}$, respectively.
We follow the 3D evolution for 218.51 s until the onset of CC.

\begin{figure*}
\includegraphics[width=0.5 \textwidth]{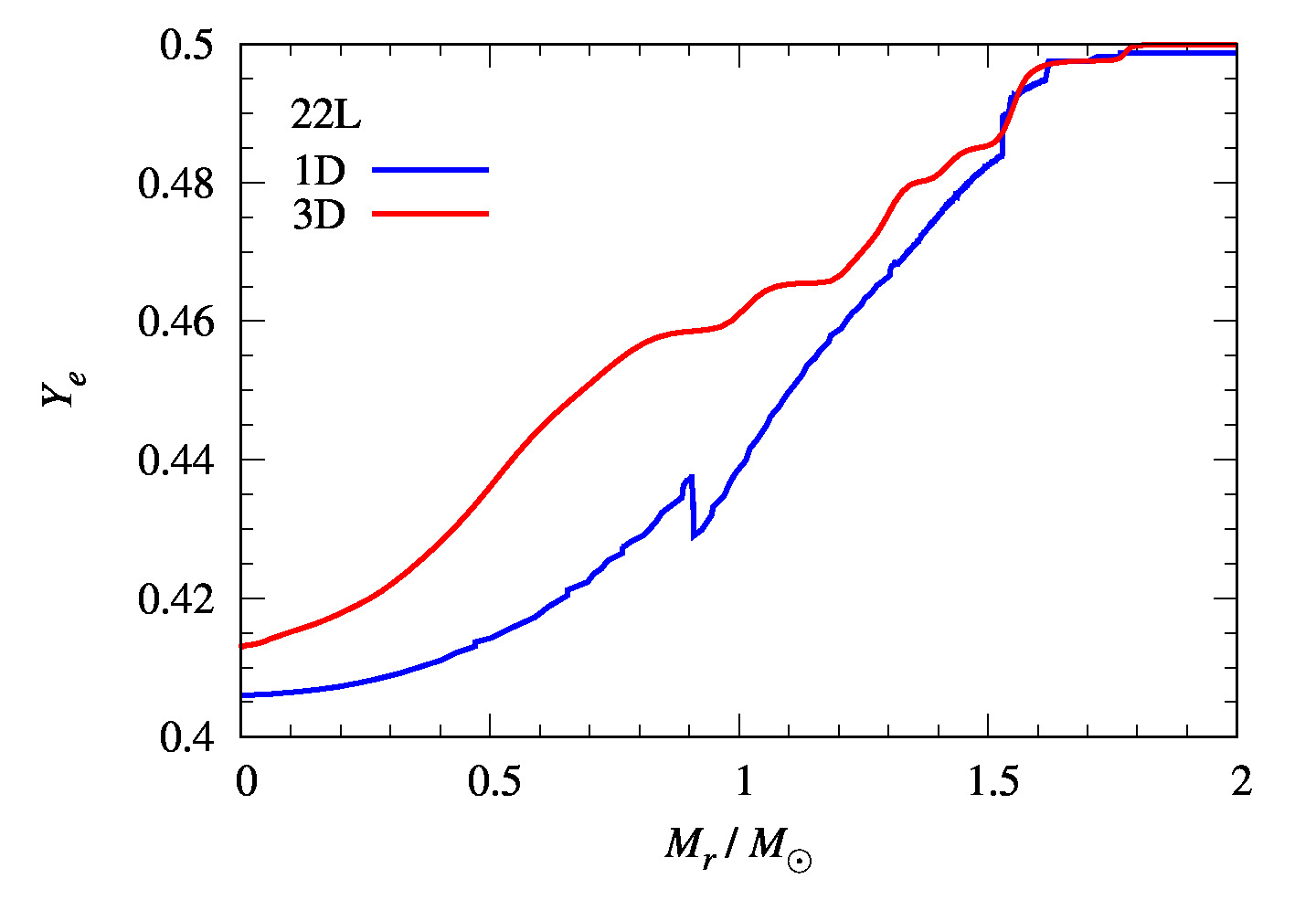}
\includegraphics[width=0.5 \textwidth]{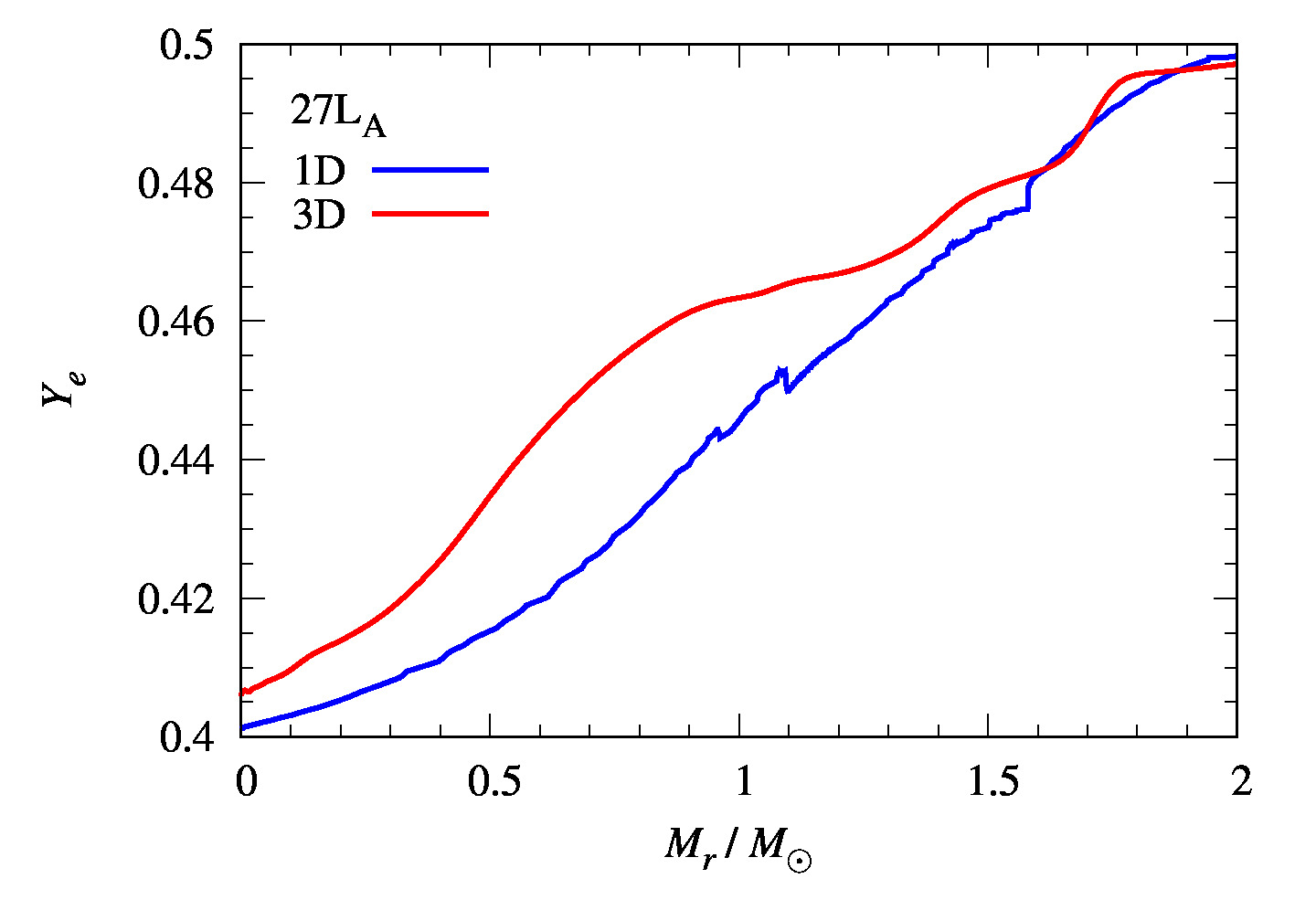}
\caption{
Electron fraction profile $Y_e$ as a function of the mass coordinate in $M_r < 2.0 M_\odot$ at the last step for models 22L (left panel) and 27\LA (right panel).
Blue and red lines are the results of 1D and 3D simulations, respectively.
\label{fig:lastye}}
\end{figure*}

\subsection{Numerical Methods} \label{subsec:hydro}
The numerical schemes in this work are the same as those in \citet{Yoshida19}. 
We employ the 3DnSEV code, which solves Newtonian hydrodynamics equations. 
The 3D models are computed on a spherical polar coordinate grid with a resolution of 
$n_r \times n_\theta \times n_{\phi} = 512 \times 64 \times 128$ zones. 
The radial grid is logarithmically spaced and covers from the center up to the outer boundary of $10^{10}$ cm.
For the polar and azimuthal angle, the grid covers all $4\pi$ sr.
For boundary conditions, reflective boundary conditions are adopted for the inner boundary.
Fixed-boundary conditions are used for the outer boundary, except for the gravitational potential that is inversely proportional to the radius at outer ghost cells.
To focus on the convective activity of the shaded regions in Figure \ref{fig:massfraction},
the inner 1000 km is solved in 1D. Gravity is included by assuming a 1D, monopole, gravitational potential. 
Such a treatment is indispensable for reducing the computational time; the nonlinear coupling between the core 
and the surrounding shells \citep[e.g.,][]{fuller15} is beyond the scope of this study.

A piecewise linear method with geometrical correction of the spherical coordinates is used to reconstruct variables at the cell interface, where a modified van Leer limiter is employed to satisfy the condition of total variation diminishing (TVD) \citep{Mignone14}.
The numerical flux is calculated by the HLLC solver \citep{Toro94}.
The Consistent Multi-fluid Advection (CMA) method of \citet{Plewa99} is used for evaluating the numerical flux of isotopes.

We use the ``Helmholtz" equation of state \citep{Timmes00}. Neutrino cooling
 is taken into account \citep{Itoh96} as a sink term in the energy equation.
As for nuclear reaction, a reaction network of 21-isotopes \citep[aprox21;][]{Paxton11} is applied when the temperature is lower than $5 \times 10^{9}$ K.
This network includes $^{54}$Fe, $^{56}$Fe, and $^{56}$Cr, which are crucial to treat a low electron fraction $Y_e \ga 0.43$ in the pre-SN stage.
Also, it is as large as that of \citet{Couch15} and a little larger than 19-isotopes of \citet{bernhard16_prog} and \citet{Yadav19}.
When the temperature is higher than $5 \times 10^{9}$ K, the chemical composition is calculated assuming nuclear statistical equilibrium (NSE) instead of solving the nuclear reaction network.

Figure \ref{fig:lastye} shows the electron fraction profile as a function of the mass coordinate in $M_r < 2.0 M_\odot$ at the last step when the central temperature reaches $10^{10}$ K for models 22L and 27\LA.
For the both models neutronization proceeds more slowly in the 3D simulations compared with the 1D simulations.
To correctly handle the neutronization of heavy elements from
Si to the iron group and the gradual shift of the nuclear abundances, one needs to use a sufficient number of isotopes \citep[$\sim$100;][]{arnett11},  which is currently computationally and technically very challenging. Since the NSE region appears mainly in the Fe core, this treatment may not significantly affect our results in which we focus on convection in the outer layers (e.g., Si/O-rich and O/Si/Ne layers).

When we start the 3D runs, seed perturbations to trigger nonspherical motions are imposed on the 1D data by introducing random perturbations of $1\%$ in density on the whole computational grid.
We terminate our 3D runs when the central temperature exceeds $10^{10}$ K, because the core is dynamically collapsing at this time.

\section{Results} \label{sec3}
In Section \ref{subsec:mach}, we start to overview the time evolution of developing turbulence in the convective layers of our 3D models.
Here we show the time evolution of the radial profiles for the angle-averaged Mach number and the turbulent velocity.
The growth of turbulence is predominantly determined by the shell burning activity, which is also mediated by the convective matter mixing there. 
To clearly show this, we present in Section \ref{sec:compos} the spatiotemporal evolution of the composition changes in association with the turbulent motions in the burning shells.
By performing a spectrum analysis, we discuss in Section \ref{subsec:powerspectra} typical scales of convective eddies in the burning shells and its relevance to the turbulent kinetic energy.
Section \ref{subsec:harmonics} is devoted to the spherical harmonics decomposition of the precollapse inhomogeneities obtained in our models, which could be applied in future SN simulations.

We note in this paper that we mainly present the results of model 27\LA rather than model 22L because the evolution of model 22L is similar to that of model 25M shown in \citet{Yoshida19}.
We performed a 3D simulation of the evolution for 105 s just before the CC of model 25M.
Model 25M has an Si/O layer of (3.0--10.5) $\times 10^{8}$ cm at the end of the simulation (when the central temperature becomes $9 \times 10^{9}$ K).
The maximum radial Mach number in this region is 0.087.
The power spectrum of the radial turbulent velocity shows the peak at $\ell = 2$.
We will show a 2D slice of the Si and O mass fractions at distinct times and the radial mass fraction distributions at the end of the simulation for model 22L in Appendix B.

\begin{figure*}
\includegraphics[width=9cm]{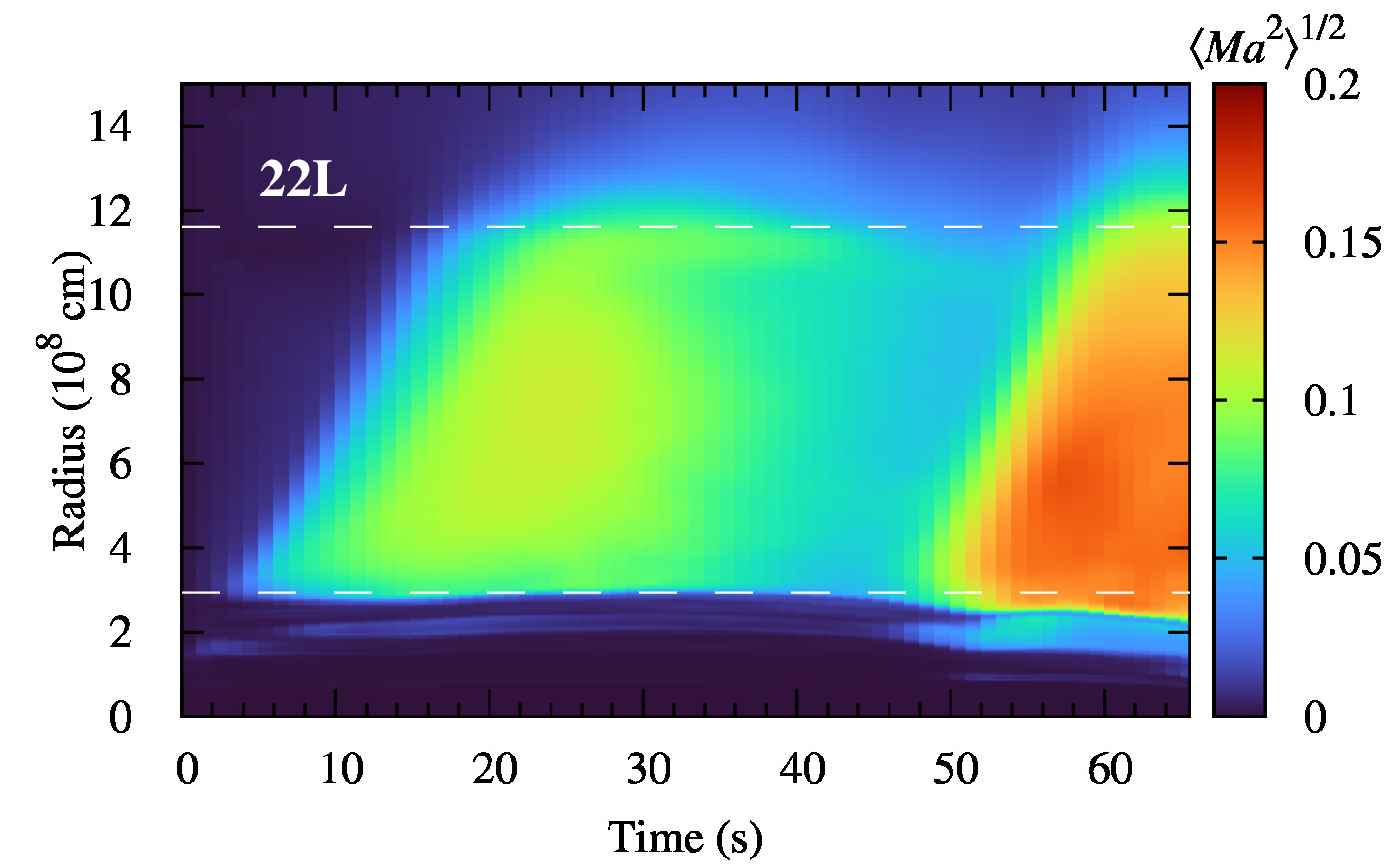}
\includegraphics[width=9cm]{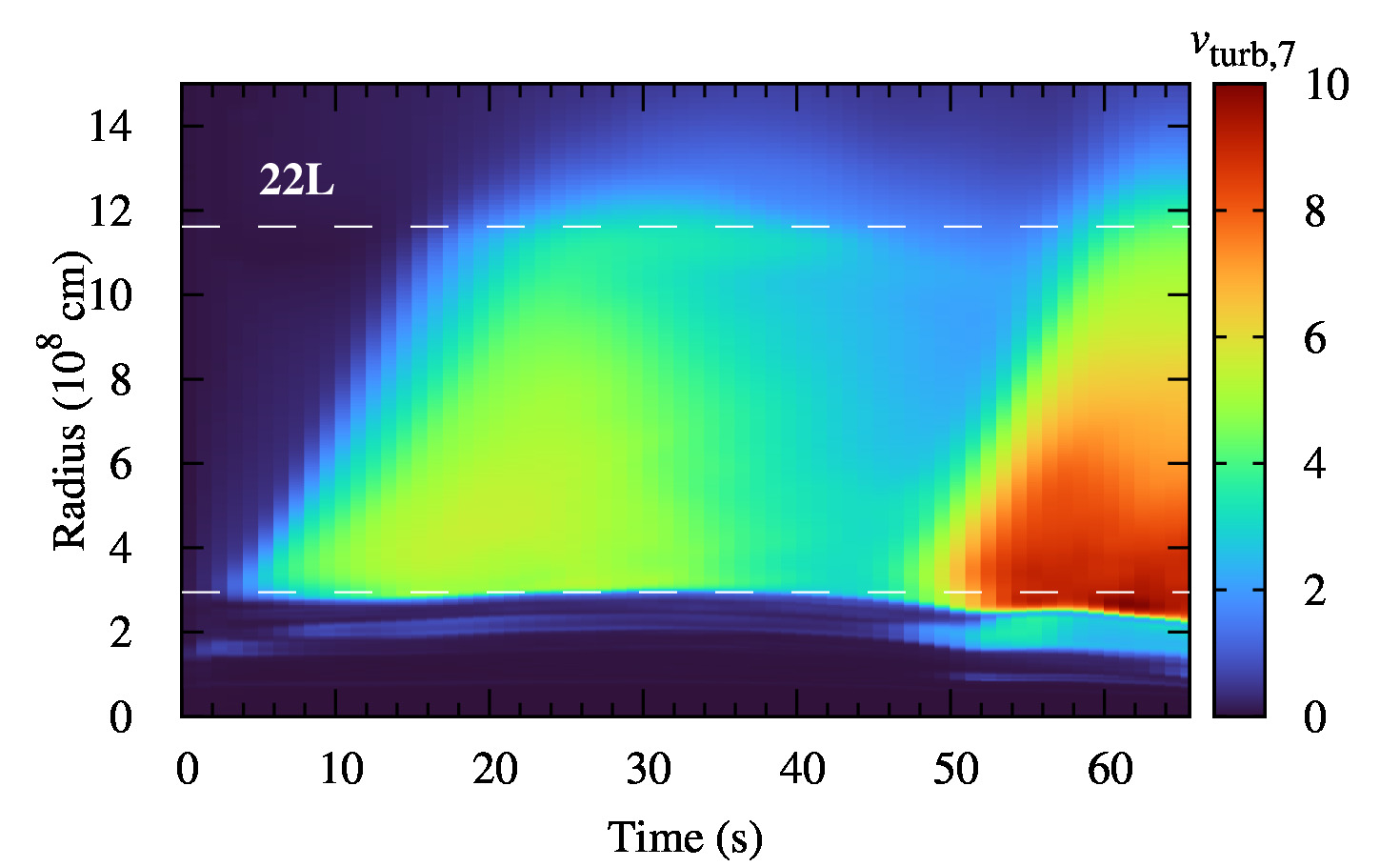}
\includegraphics[width=9cm]{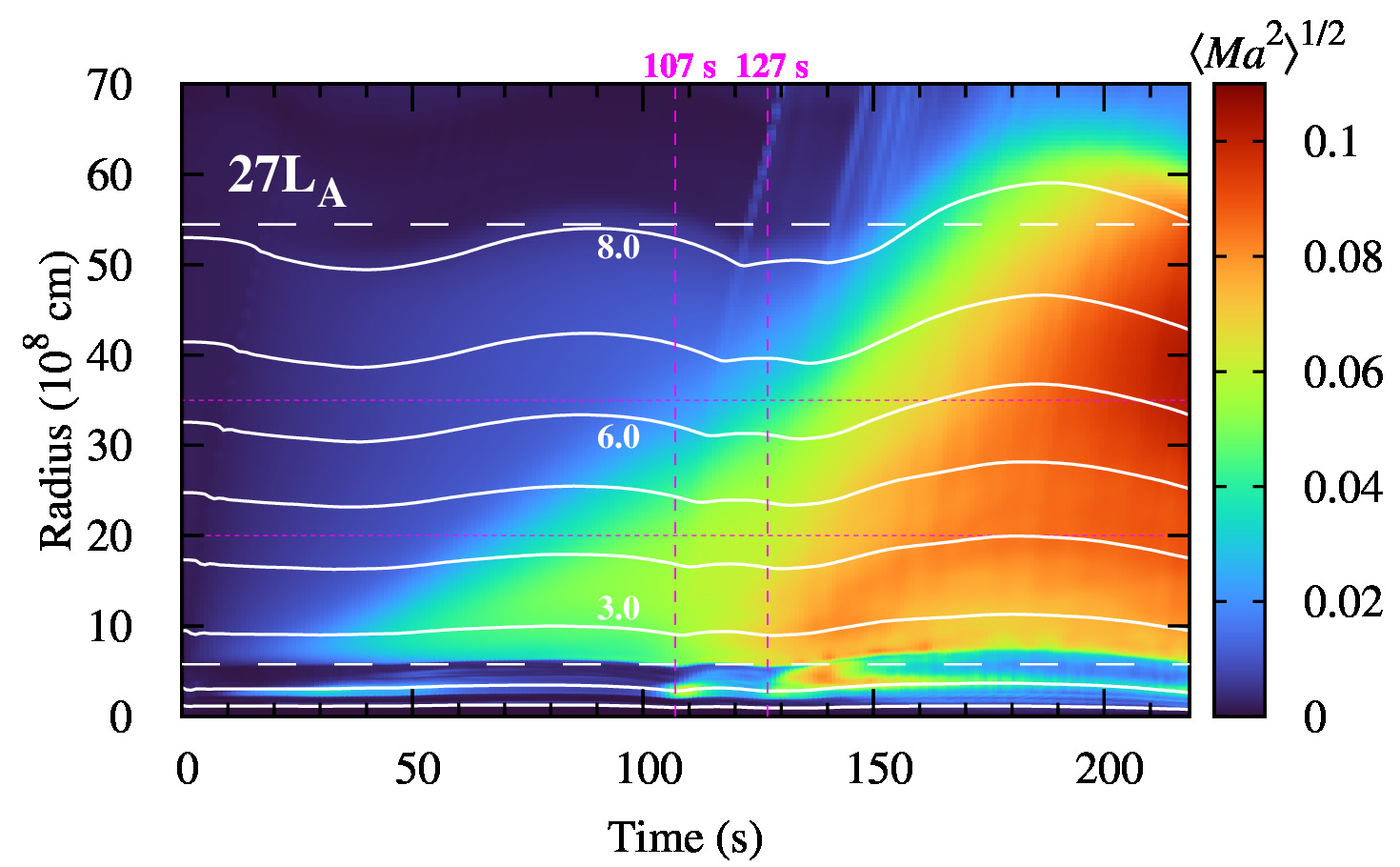}
\includegraphics[width=9cm]{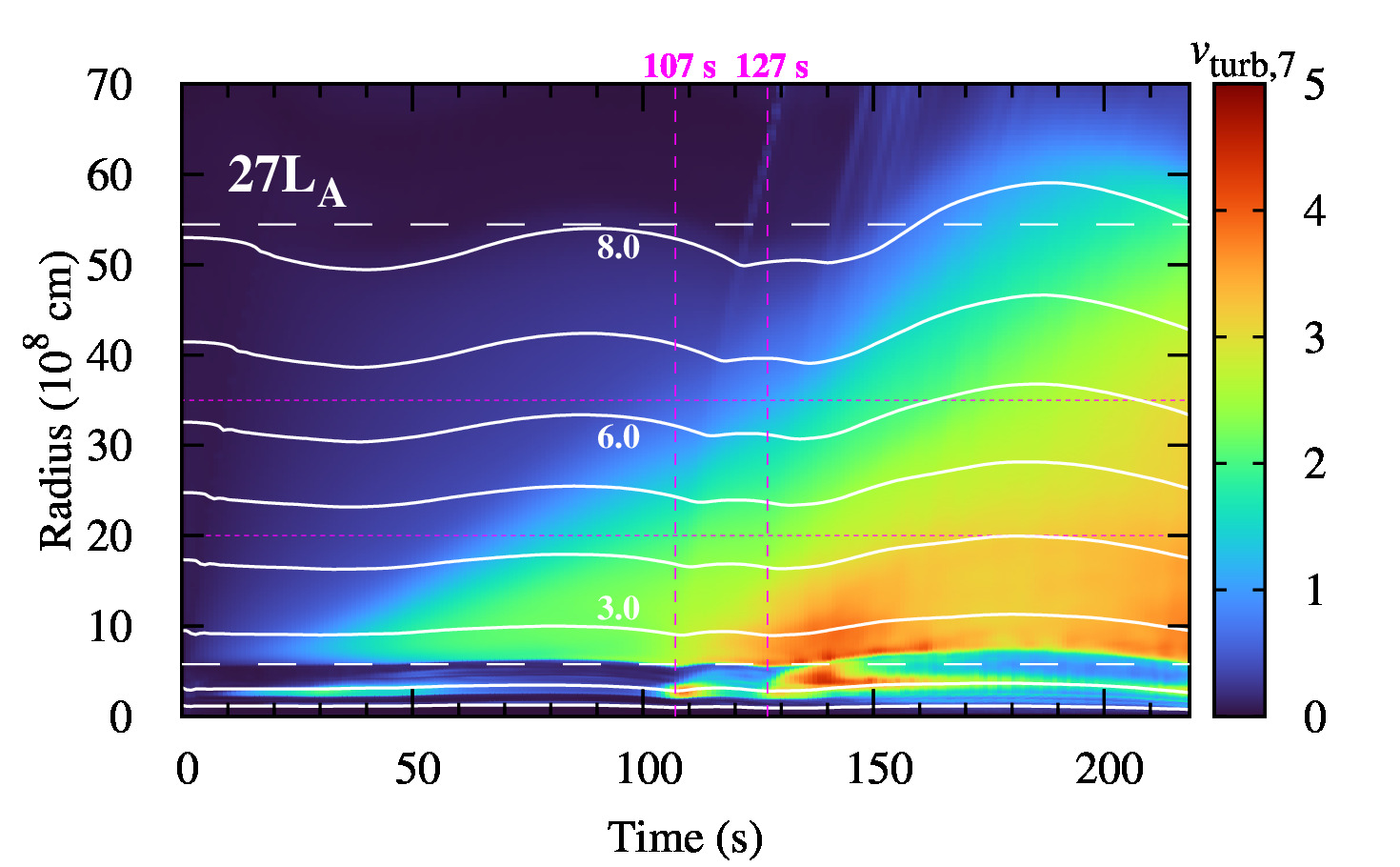}
\caption{Spatiotemporal evolution of the angle-averaged turbulent Mach number ($\langle \Mach^2 \rangle^{1/2}(r)$; top panels) and angle-averaged turbulent velocity in units of $10^{7}$ cm s$^{-1}$ ($v_{{\rm turb},7}$; bottom panels).
The left and right panels are models 22L and 27\LA, respectively.
In both of the panels, the region sandwiched by the  horizontal white dashed lines corresponds to the green hatched region (convective) in Figure \ref{fig:massfraction}, namely,
 the inner and outer boundary of the Si/O layer (left panel) and O/Si/Ne layer (right panel).
In the right panel, the two vertical magenta lines  denote the beginning of the phase II ($t = 107$ s) and phase III ($t = 127$ s) (see text for details).
\label{fig:Ma}}
\end{figure*}
\begin{figure}
\includegraphics[width=\columnwidth]{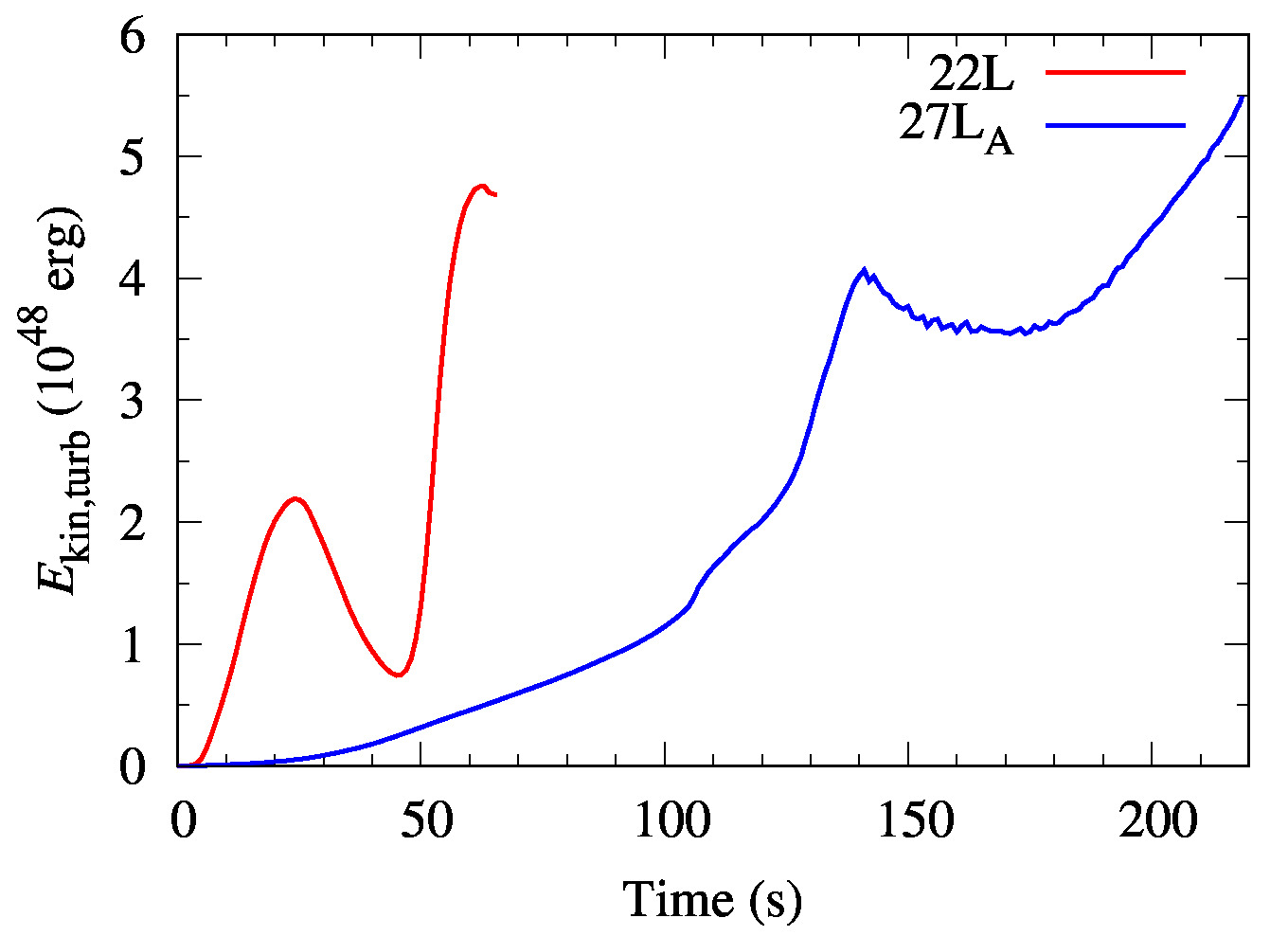}
\caption{Time evolution of the turbulent kinetic energy $E_{\rm kin,dis}$ in the Si- and O-rich region.
Red and blue curves indicate models 22L and 27\LA, respectively.
The inner and outer boundaries of the Si and O-rich region are the radii where the O and Si mass fractions are less than 0.01, respectively.
\label{fig:Ekindis}}
\end{figure}

\subsection{Turbulent Mach Number} \label{subsec:mach}
Figure \ref{fig:Ma} shows the spatiotemporal evolution of the angle-averaged turbulent Mach number (left panels, $\langle \Mach^2 \rangle^{1/2}(r)$) and the turbulent velocity in units of $10^{7}$ cm s$^{-1}$ (right panels, $v_{\rm turb,7}$) for models 22L (top panels)  and 27L$_{\rm A}$ (bottom panels), respectively. 
Here the definition of $\langle \Mach^2 \rangle^{1/2}(r)$ is the same as in Equation (8) in \citet{Yoshida19}, 
\begin{equation}
\langle \Mach^2 \rangle^{1/2}(r) = \left[ \frac{\int \rho \{(v_r - \langle v_r \rangle)^2 + v_\theta^2 + v_\phi^2\}d\Omega}{\int \rho c_s^2\,d\Omega} \right]^{1/2},
\end{equation}
where $\rho$ is the density; $v_r$, $v_\theta$, and $v_\phi$ are radial, tangential, and azimuthal velocities, respectively; $\langle v_r \rangle$ is the angle-averaged radial component velocity obtained by Reynolds averaging; $\Omega$ is the solid angle; and $c_s$ are the sound speed.
The definition of the turbulent velocity $v_{\rm turb}$ is defined as
\begin{equation}
    v_{\rm turb} = \left[ \frac{\int \rho \{(v_r - \langle v_r \rangle)^2 + v_\theta^2 + v_\phi^2\}d\Omega}{\int \rho \,d\Omega} \right]^{1/2}.
\end{equation}
Figure \ref{fig:Ekindis} is the time evolution of the turbulent kinetic energy $E_{\rm kin,turb}$ in the Si- and O-rich region defined as
\begin{equation}
    E_{\rm kin,turb} = \int \frac{1}{2} \rho \{(v_r - \langle v_r \rangle)^2 + v_\theta^2 + v_\phi^2\} r^{2}dr d\Omega.
\end{equation}
The inner and outer boundaries of the Si- and O-rich region are determined by the conditions of the O mass fraction larger than 0.01 and the Si mass fraction larger than 0.1, respectively.
Note that we use the term turbulent velocity  as a velocity fluctuation in this paper even when the fluctuation has not developed to turbulence.

The top panels of Figure \ref{fig:Ma} show that in model 22L turbulence starts to develop from $t = 10$ s at the base of the Si/O layer ($r \sim 3 \times 10^{8}$ cm), which is driven by O shell burning. 
Note that $t = 0$ is defined as the epoch of the start of the 3D simulation. 
As seen, turbulent flows with the Mach number of $\sim$0.1 (greenish region) spread over the entire Si/O layer (up to the upper white dashed line) before the final simulation time of this model, $t = 65.5$ s (before CC onset). 
At $t \sim 50$ s, a strong turbulence of the Mach number $\sim$0.16 and the turbulent velocity $\sim$10$^{8}$ cm s$^{-1}$ develops transiently from the base of the Si/O layer (again triggered by O shell burning).
Figure \ref{fig:Ekindis} shows that the kinetic energy of the turbulent motion increases after $\sim$50 s.
At this time, the nuclear energy generation rate at the bottom of this convective layer reaches $\sim$6$\times 10^{16}$ erg g$^{-1}$ s$^{-1}$, whereas it is less than $1 \times 10^{16}$ erg g$^{-1}$ s$^{-1}$ before that time.
This kind of episodic burning causes a slight expansion and the subsequent contraction of the above layer, which was already found in 3D models of 25M \citep{Yoshida19} and also of the $18.8 M_{\odot}$ star by \citet{Yadav19}.
In Figure \ref{fig:vrturb_xz} in Appendix B, we show 2D slices on the \textit{x-z} plane of the radial turbulent velocity distributions at 10, 40, 50, and 65.5 s in model 22L.

Note the difference of the radial scale between the 
top and bottom panels of Figure \ref{fig:Ma}.
The turbulent flows develop in the more extended region for model 27L$_{\rm A}$, whereas the maximum Mach number ($\sim$0.1) is smaller than that of model 22L (compare the left and right panels). 
In model 27\LA, the temperature at the bottom of the convective layer is lower and the neon shell burning occurs.
The nuclear generation rate by the neon shell burning (model 27\LA) is much smaller than the oxygen shell burning (model 22L).
Given the initial composition profile (i.e., the broader shaded region; see Figure \ref{fig:massfraction}), this may not be very surprising.
However, such a large-scale mixing has been first obtained in our model series. 
Hence, we pay particular attention to this model.
We show 2D slices on the \textit{x-z} plane for the radial turbulent velocity distributions at 10, 90, 140, and 218.5 s for model 27\LA in Figure \ref{fig:vrturb_xz} in Appendix B.

For better understanding, we divide the entire evolution of model 27L$_{\rm A}$ into the three phases, namely, 
phase I (0 s $\leq t \leq 107$ s), phase II (107 s $\leq t \leq 127$ s), and phase III  (127 s $\leq t \leq 218$ s), respectively. 
In the phase I, turbulence starts from the base of the O/Si/Ne layer 
($r \sim 6 \times 10^{8}$ cm at $t = 10$ s), which gradually extends outward. 
In the phases II and III, the more enhanced turbulence stems, the maximum angle-averaged turbulent velocity is $4.2 \times 10^{7}$ cm s$^{-1}$, from stronger (Ne) shell burning.
We explain more in detail in the next section.

\subsection{Composition Distribution} \label{sec:compos}

\begin{figure}
\includegraphics[width=\columnwidth]{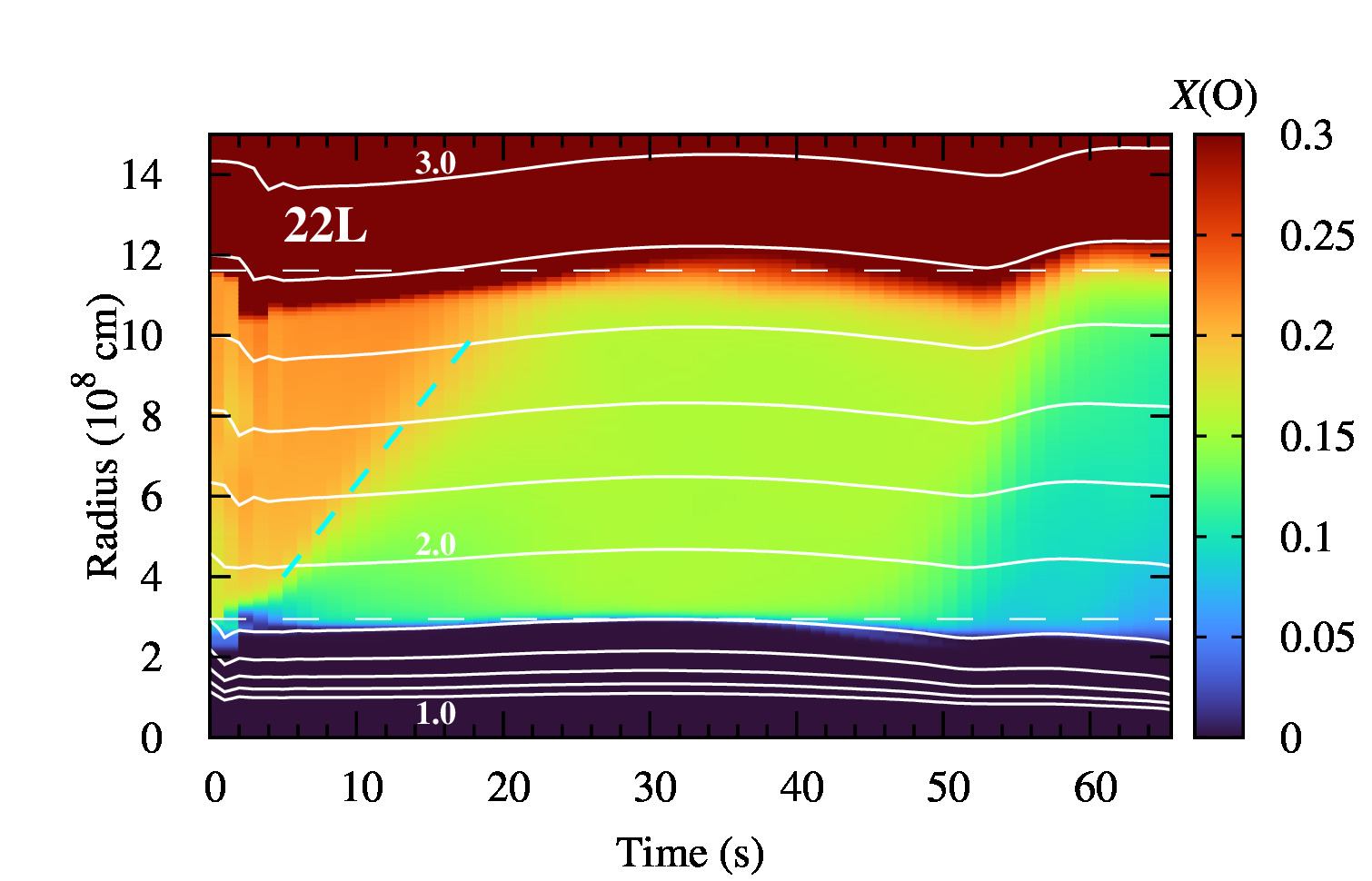}
\caption{Sspatiotemporal evolution of the angle-averaged O mass fraction for model 22L.
The white dashed lines indicate the initial radii of the inner and outer boundaries of the Si/O layer.
The white curves indicate the mass coordinates from 1.0 to 3.0 $M_\odot$ in intervals of 0.2 $M_\odot$.
The cyan dashed line corresponds to the extension of oxygen depletion for the first $\sim$20 s.
\label{fig:mforav22L}}
\end{figure}

\begin{figure}
\includegraphics[width=\columnwidth]{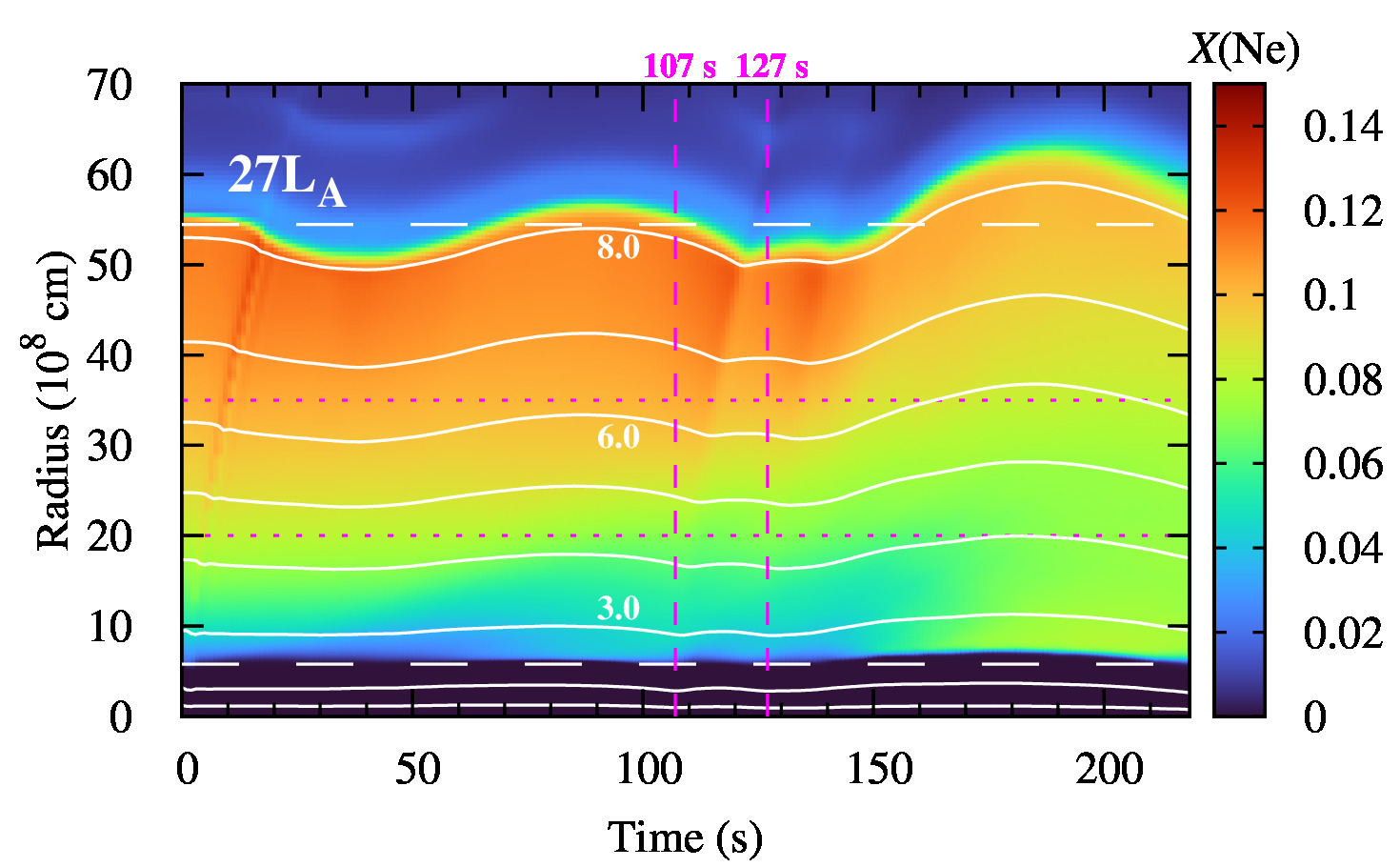}
\includegraphics[width=\columnwidth]{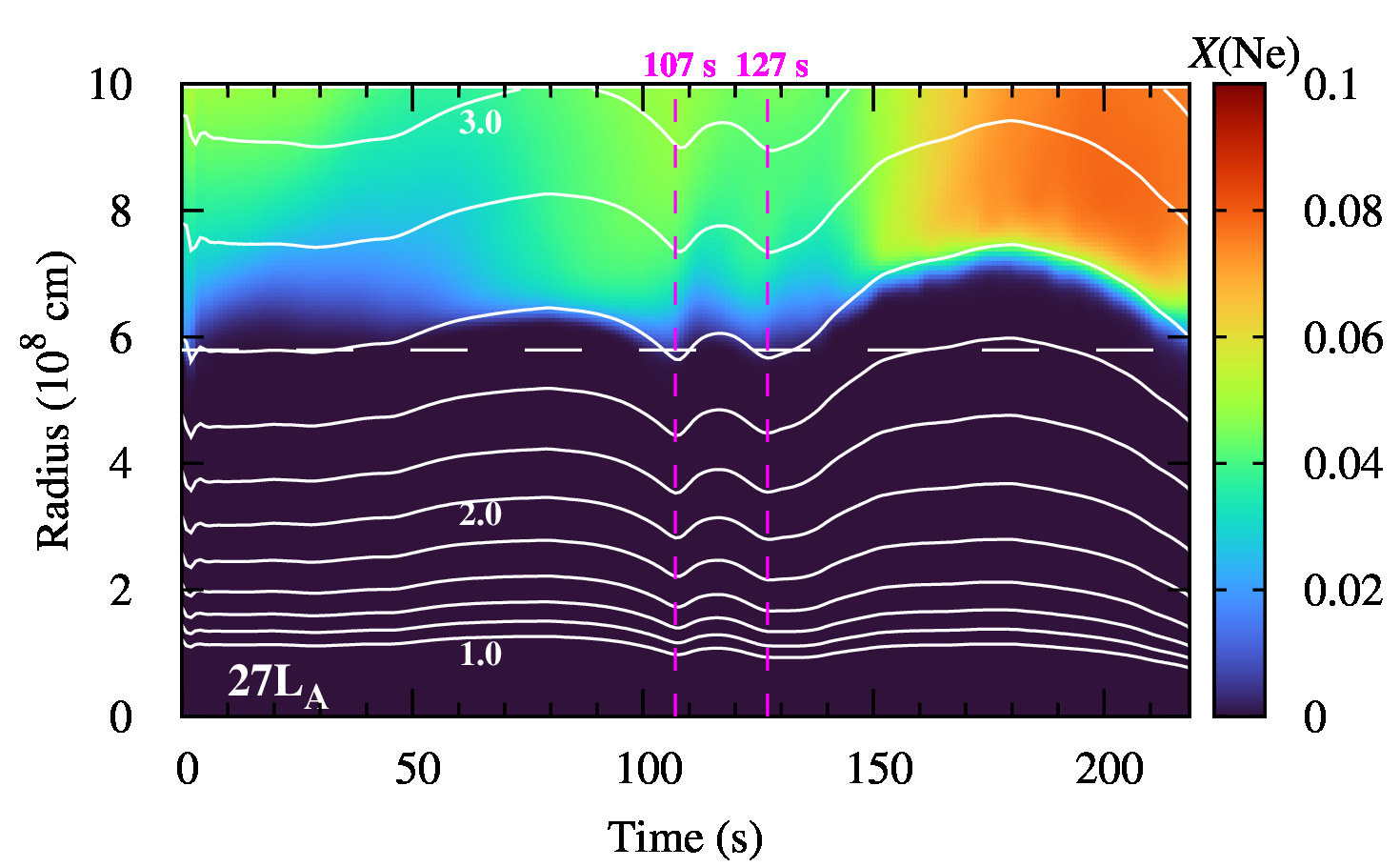}
\caption{Spatiotemporal evolution of the angle-averaged Ne for model 27\LA.
The top panel includes the whole convective (O/Si/Ne) layer.
The white curves indicate the mass coordinates in intervals of 1 $M_\odot$.
The bottom panel is focusing on the region close to the lower boundary of the convective (O/Si/Ne) layer 
(note that the location of the horizontal dashed line is at 5.8 $\times 10^{8}$ cm). 
The white curves indicate the mass coordinates from 1.0 to 3.0 $M_\odot$ in intervals of 0.2 $M_\odot$.
The two magenta vertical lines correspond to the beginning of phases II ($t = 107$ s) and III ($t = 127$ s).
\label{fig:mfrav27LA_Ne}}
\end{figure}

\begin{figure}
\includegraphics[width=\columnwidth]{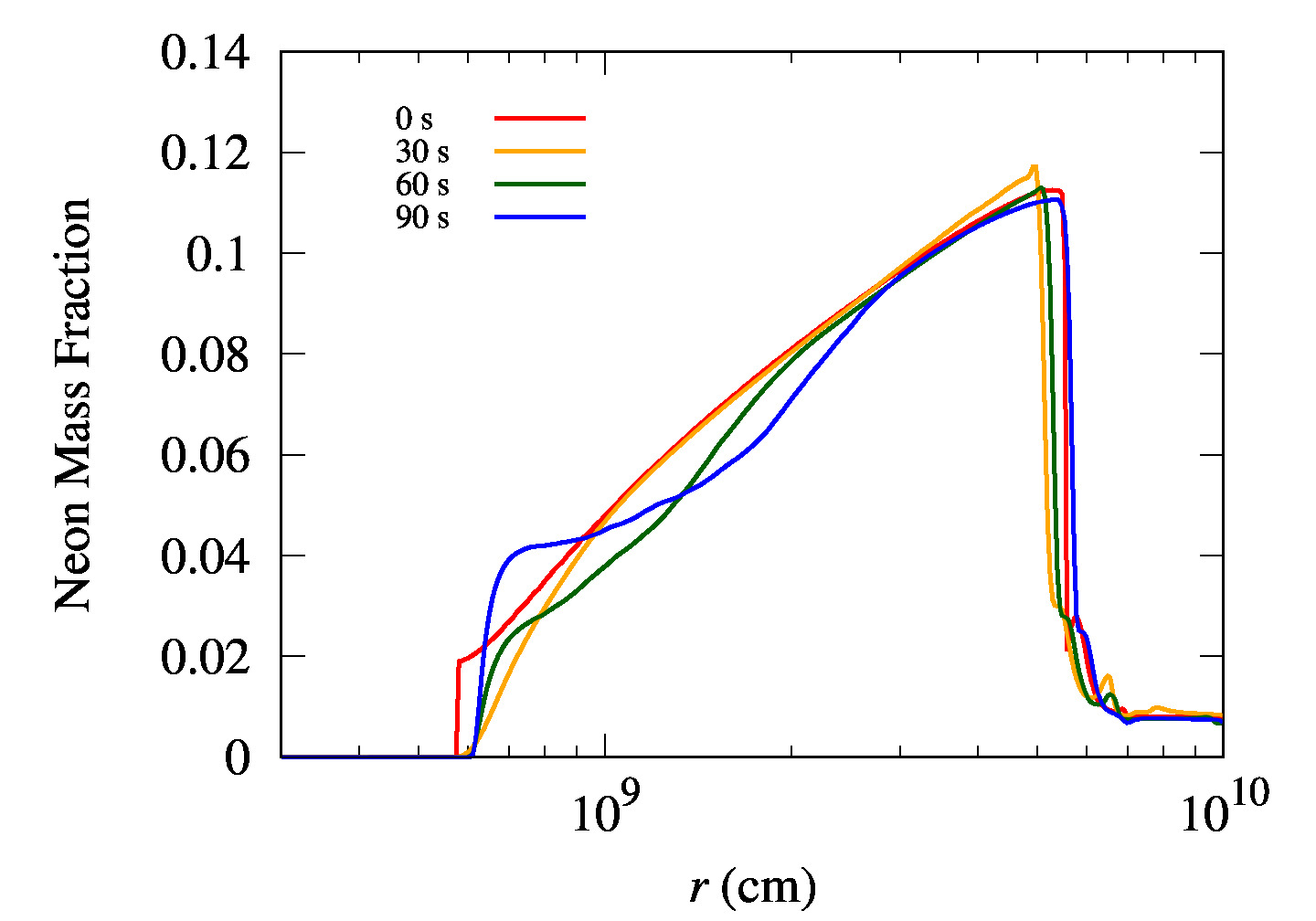}
\includegraphics[width=\columnwidth]{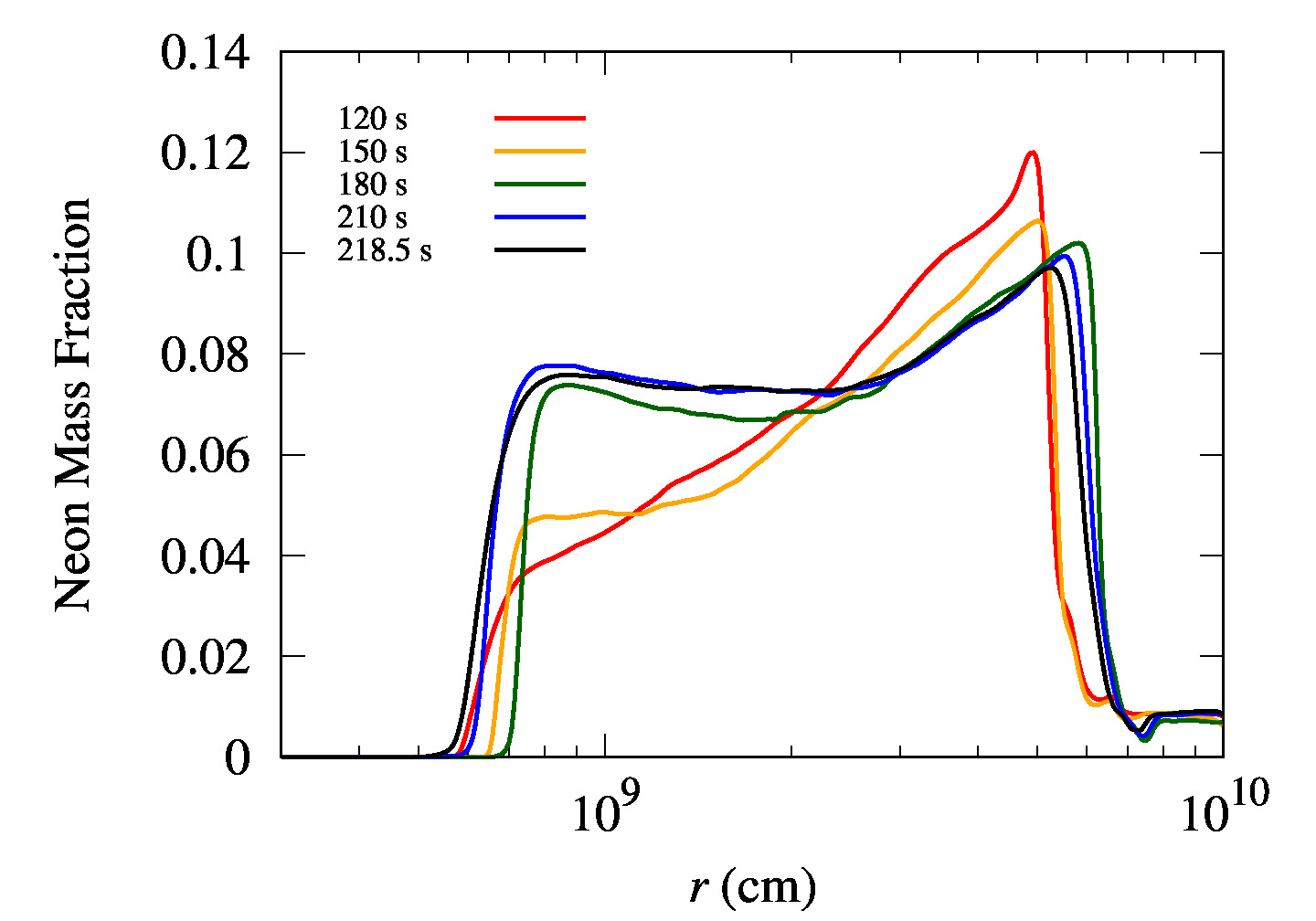}
\caption{Radial profile of neon mass fraction in model 27\LA.
The top panel indicates the profile at $t =$ 0 (red), 30 (orange), 60 (green), and 90 s (blue).
The bottom panel indicates the profile at $t =$ 120 (red), 150 (orange), 180 (green), 210 (blue), and 218.5 s (black).
\label{fig:mfner27LA}}
\end{figure}

\begin{figure*}
\includegraphics[width=9cm]{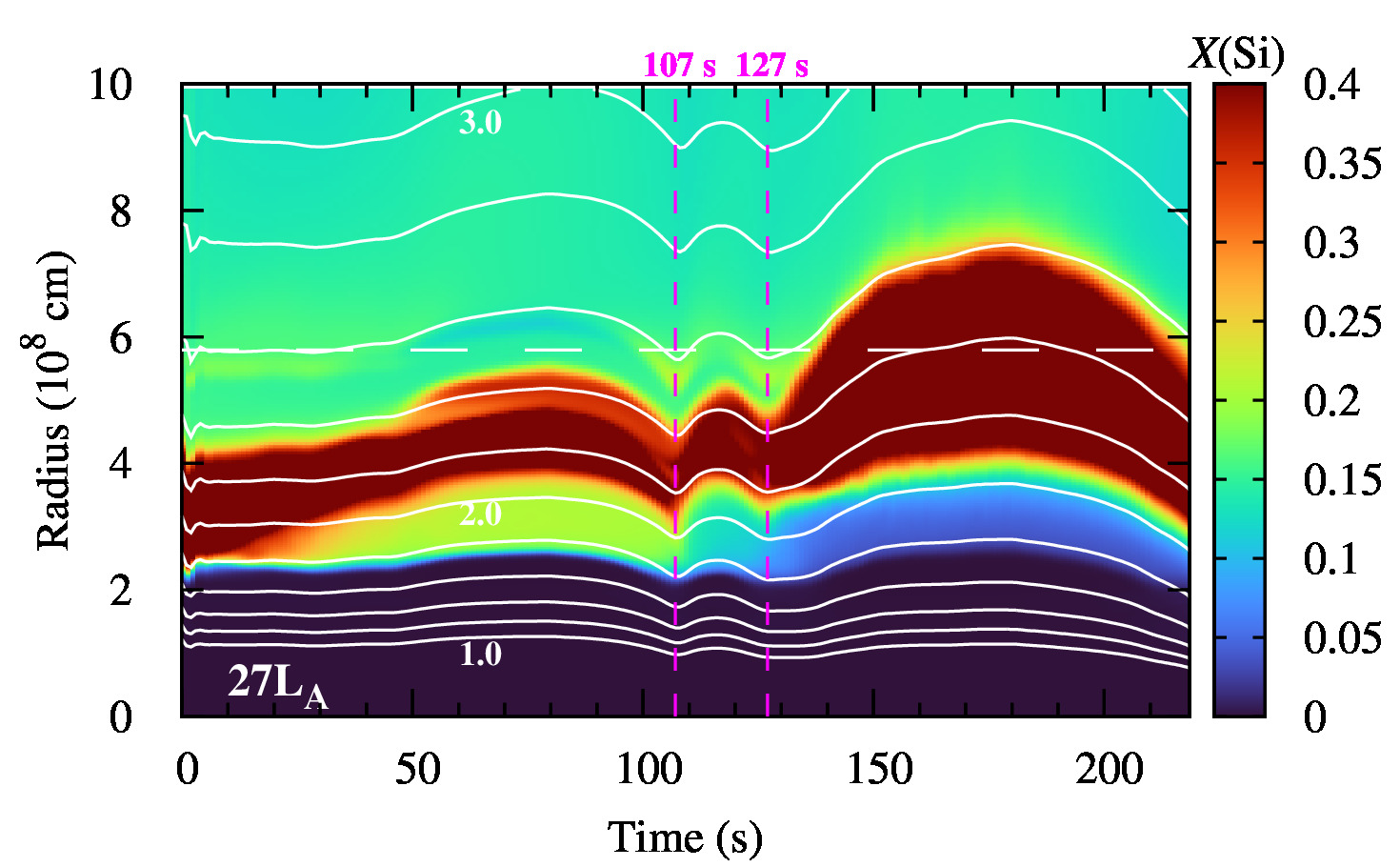}
\includegraphics[width=9cm]{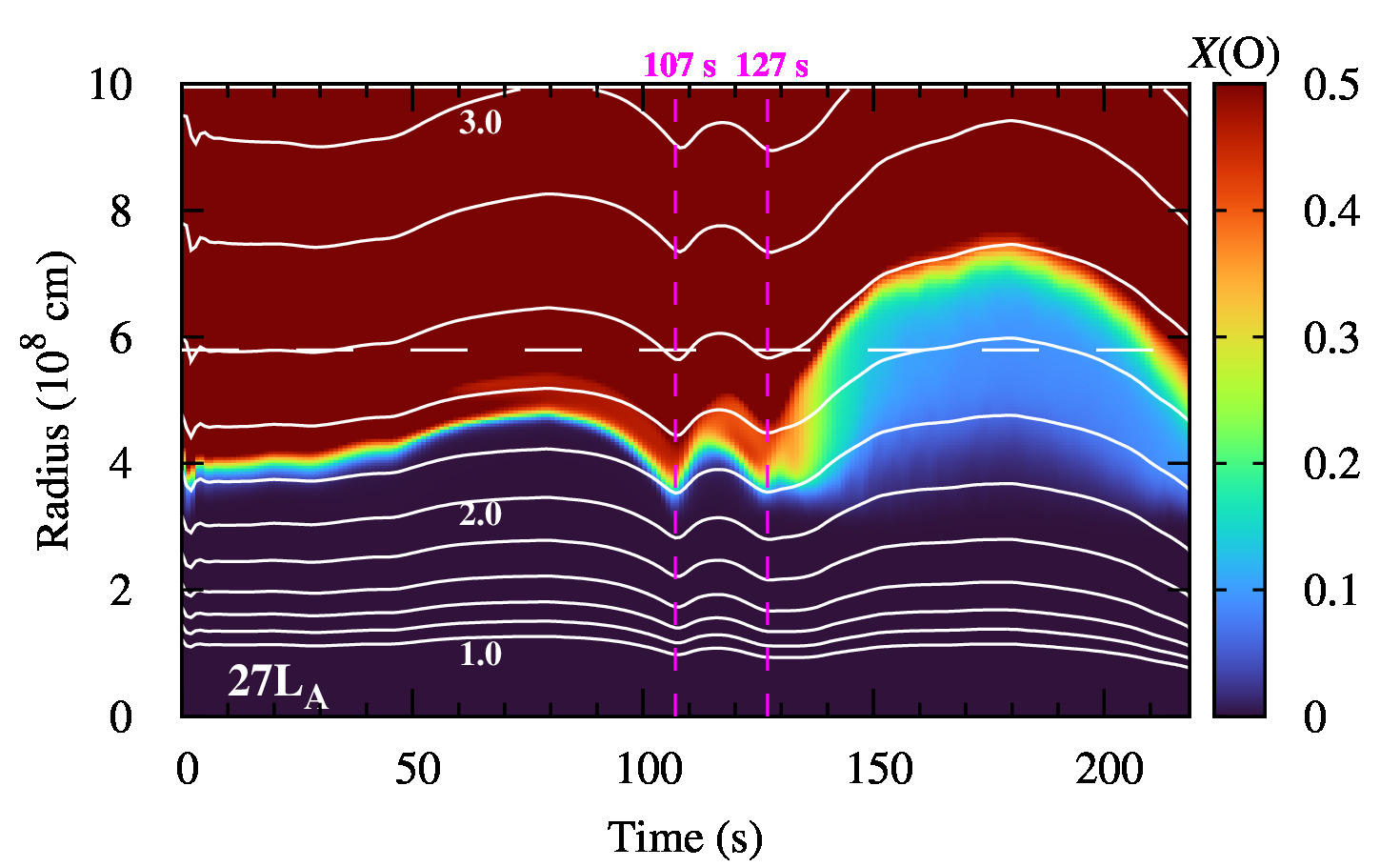}
\caption{Spatiotemporal evolution of the angle-averaged mass fractions of Si (left panel) and O (right panel) for model 27\LA, focusing on the region close to the lower boundary of the convective (O/Si/Ne) layer (note that the location of the horizontal dashed line is at 5.8 $\times 10^{8}$ cm).
The white curves indicate the mass coordinates from 1.0 to 3.0 $M_\odot$ in intervals of 0.2 $M_\odot$.
The two magenta vertical lines correspond to the beginning of phases II ($t = 107$ s) and III ($t = 127$ s).
\label{fig:mfrav27LA_SiOinner}}
\end{figure*}

We move on to explore how the turbulent activity shown in Figure \ref{fig:Ma} is triggered in association with the change in the element composition in the burning shells.

\subsubsection{Model 22L}

First, let us focus on the mass fraction distribution for model 22L.
Figure \ref{fig:mforav22L} shows the spatiotemporal evolution of the angle-averaged O mass fraction for model 22L.
One sees that shortly after the start of the calculation ($t = 0$), the oxygen-depleted materials are mixed into the Si/O layer (0 s $\leq t \leq 20$ s, shown as a region below the cyan line).
Subsequently, the turbulent mixing encompasses the entire Si/O layer ($3 \times 10^{8}$ cm $\leq r \leq 12 \times 10^{8}$ cm, corresponding to the shaded region in the left panel of Figure 1) up to $t \sim 35$ s with the diminishing turbulent activity (see the top panels of Figure \ref{fig:Ma}). 
After $t \sim$ 50 s, the O mass fraction decreases at the bottom of the Si/O layer owing to (the episodic) O shell burning, and the turbulent flows radially growing outward result in the reduction of the O mass fraction in the outer layer (see the bluish region for 50 s $\leq t \leq 60$ s and Figure \ref{fig:2Dmfo22L}).

\begin{figure}
\includegraphics[width=\columnwidth]{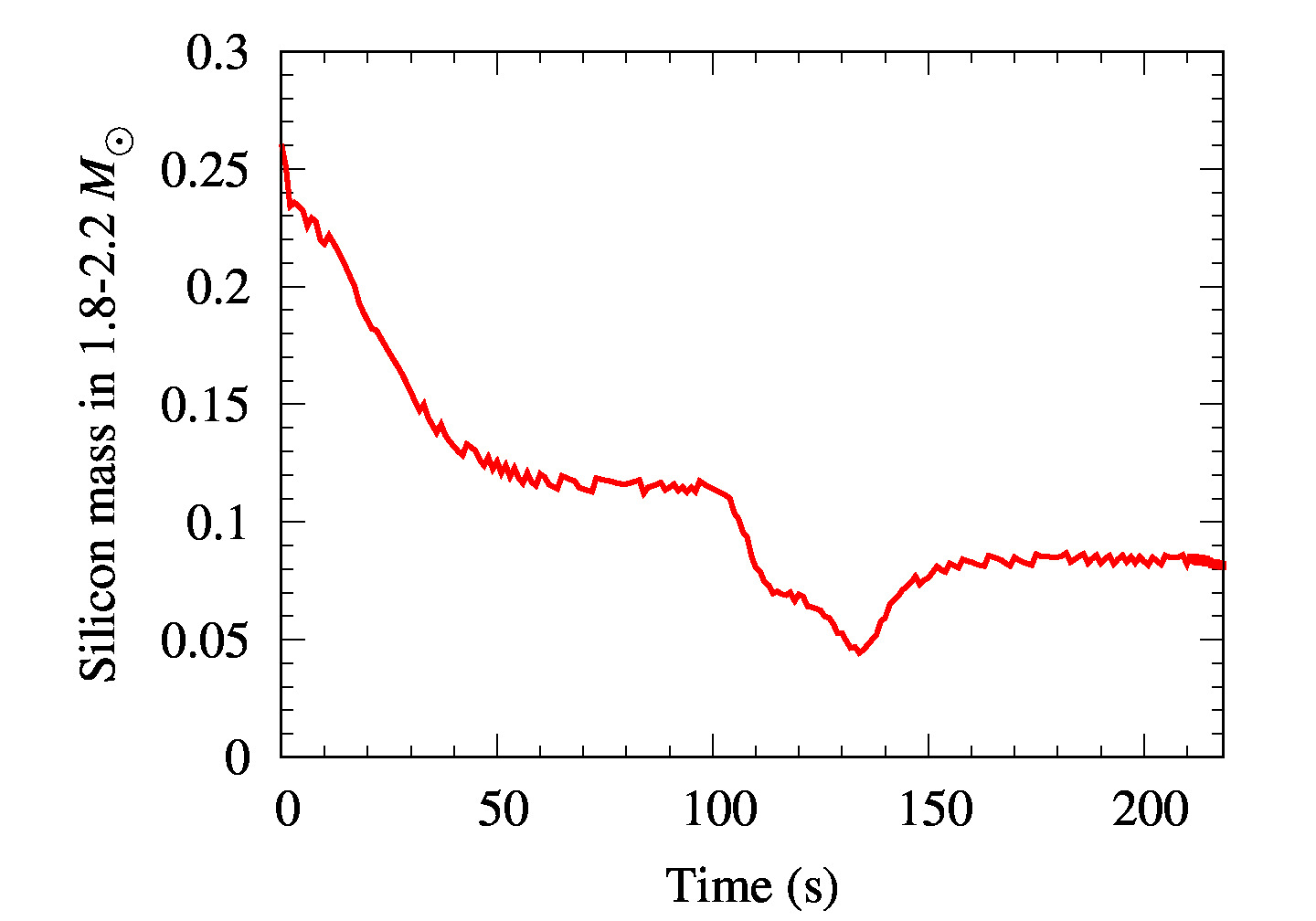}
\caption{Time evolution of the total Si mass in the region of 1.8--2.2 $M_\odot$ in model 27\LA.
\label{fig:simass27LA}}
\end{figure}

\begin{figure}
\includegraphics[width=\columnwidth]{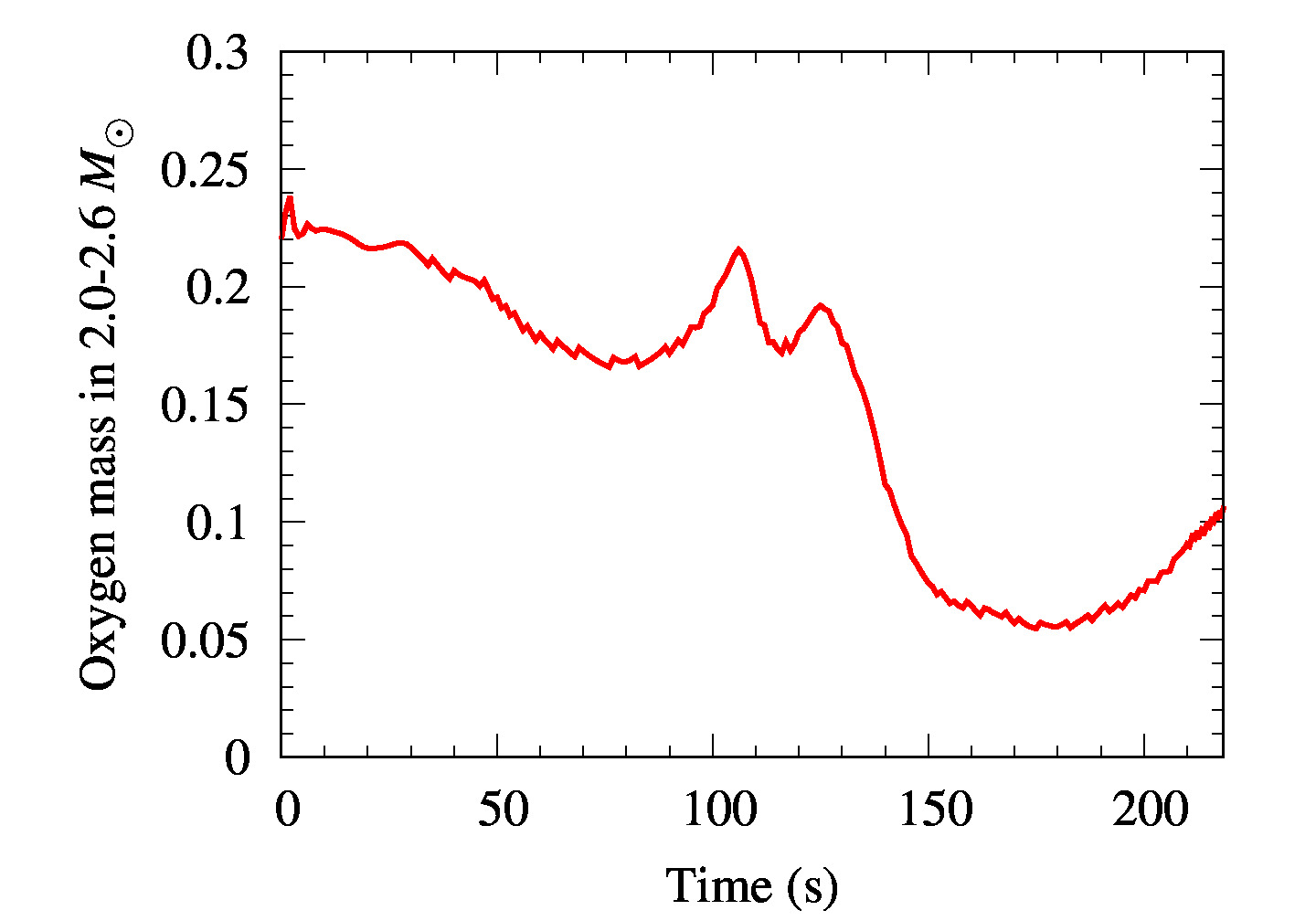}
\caption{Time evolution of the total O mass in the region of 2.0--2.6 $M_\odot$ in model 27\LA.
\label{fig:omass27LA}}
\end{figure}

\begin{figure}
\includegraphics[width=\columnwidth]{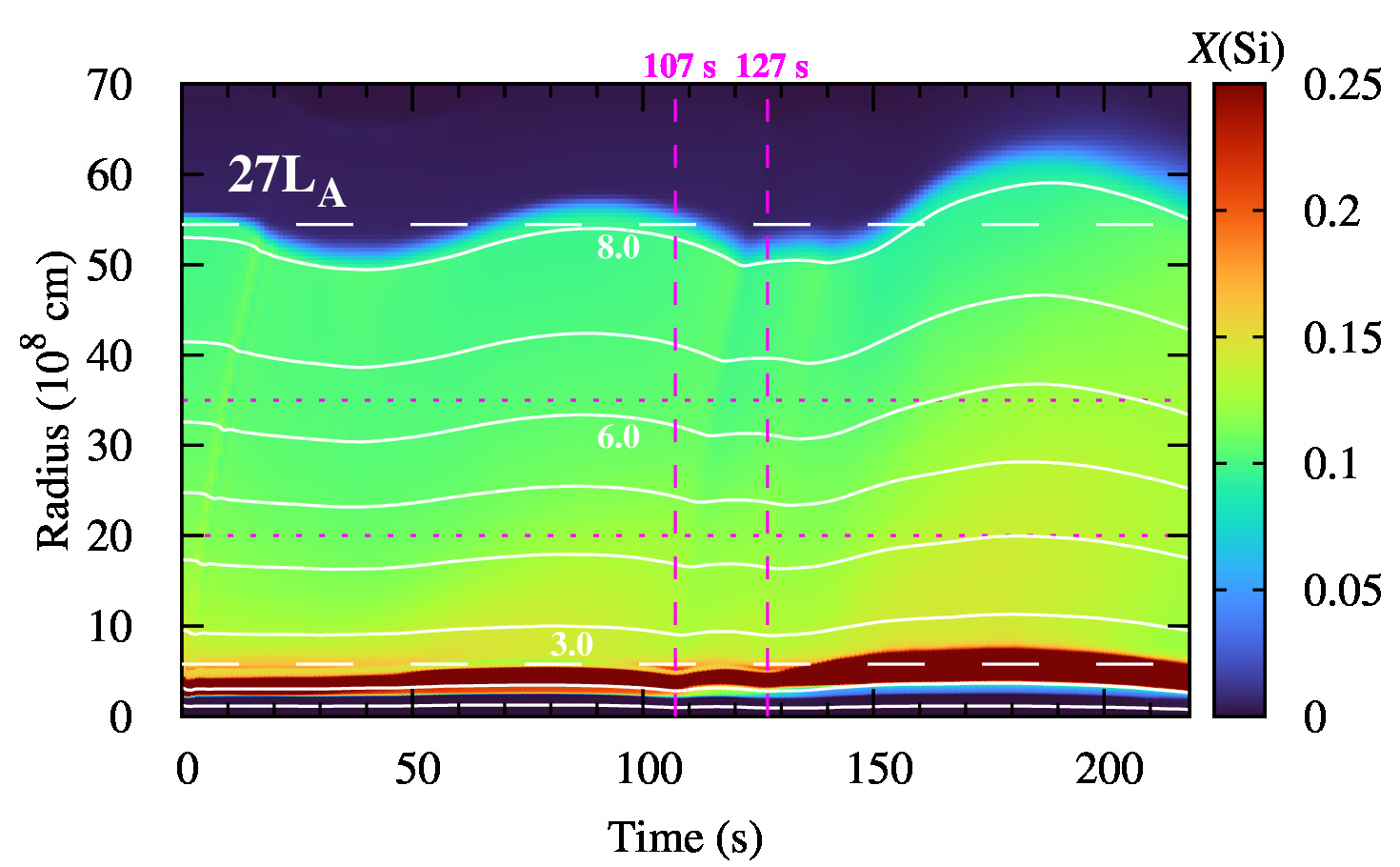}
\caption{
Spatiotemporal evolution of the angle-averaged Si mass fraction in the whole convective O/Si/Ne layer for model 27\LA.
The white curves indicate the mass coordinates from 1.0 to 8.0 $M_\odot$ in intervals of 1 $M_\odot$.
The white dashed lines indicate the initial radii of the inner and outer boundaries of the O/Si/Ne layer.
The two magenta horizontal lines denote the radii of $2.0 \times 10^{9}$ cm and $3.0 \times 10^{9}$ cm.
The two magenta vertical lines correspond to the beginning of phases II ($t = 107$ s) and III ($t = 127$ s).}
\label{fig:mfrav27LA}
\end{figure}

\subsubsection{Model 27\LA}

Second, we will show the mass fraction distributions of O, Ne, and Si in model 27\LA.
We discuss shell burning processes in the Si/O-rich and O/Si/Ne layers and the development of the turbulent motion in these layers.

\paragraph{Ne mass fraction}
Figure \ref{fig:mfrav27LA_Ne} shows the spatiotemporal evolution of the Ne mass fraction.
The bottom panel is focusing on the region close to the lower boundary of the O/Si/Ne layer.
Comparing with the bottom right panel of Figure \ref{fig:Ma}, a cautious reader would notice that the reduction of the Ne mass fraction is associated with the high turbulent velocity in this layer.
We will describe details below.

In the phase I (0 s $\leq t \leq 107$ s), the Ne burning near the base of the layer close to the lower horizontal line results in the reduction of the Ne mass fraction in the region $r \la 1.5 \times 10^{9}$ cm.
Looking at the same region to the right up to $t \sim$ 50--100 s, one can see that the radial gradient of the Ne mass fraction becomes less steep owing to turbulent matter mixing, by which the Ne-rich material is mixed downward via downflows into the Ne-poor region. 
We also see the above trend in the top panel of Figure \ref{fig:mfner27LA}, which indicates the radial profiles of the angle-averaged Ne mass fraction at definite times with intervals of 30 s.
The growth of the turbulent region from $r \sim 7 \times 10^{8}$ cm at $t \sim 30$ s can also be depicted in the spacetime diagram of the turbulent velocity, which continues into the phase III up to the radius of the outer boundary of the layer at $r \sim 60 \times 10^{8}$ cm.
In the phase II and also in the phase III, the growth of high turbulent activity is observed as a red region above the shell (the lower horizontal white dashed line). 
This is again predominantly because of the Ne shell burning there. 
The reduction of the Ne mass fraction is seen in the regions $r \la 1 \times 10^{9}$ cm and $r \sim 4 \times 10^{9}$ cm around $t \sim 150$ s in the top panel.  
Besides, one can see the episodic burning below the base of the O-rich layer (see the two red regions near at $t \sim 100$ s and 127 s $\leq t \la 150$ s below the lower horizontal white dashed line). 
This plays a key role for enhancing the turbulent mixing as we will explain in what follows.

\paragraph{Silicon and oxygen mass fractions in the region close to the inner boundary of the O/Si/Ne layer}
Figure \ref{fig:mfrav27LA_SiOinner} is for the Si (left panel) and O (right panel) mass fraction of model 27\LA, focusing on the region close to the inner boundary of the O/Si/Ne layer (note the location of the horizontal white dashed line at $5.8 \times 10^8$ cm).
By these zoom-in plots, we focus on why the episodic expansion occurs, which is seen as the three bumps in the red regions in the phases I, II, and III in the left panel.

From the left panel of Figure \ref{fig:mfrav27LA_SiOinner}, one can see that the gradual decrease of the Si mass fraction takes place at $2 \times 10^8$ cm $\la r \la 3 \times 10^8$ cm, $M_r \sim 2$ $M_{\odot}$, at 0 s $\leq t \la 107$ s in the phase I.
Figure \ref{fig:simass27LA} shows that the Si mass in this region (1.8 $M_\odot \la M_r \la 2.2 M_\odot$) gradually decreases for the former $\sim$50 s.
This gradual Si shell burning causes the first bump, which can be seen as an upward lift of the Si-rich material up to $M_r \lesssim 2.6 M_{\odot}$. 
Looking at the same region to the right (namely, phase II), one can see the abrupt decrease of the Si mass fraction (the color changes from orange to blue/green), which is due to the episodic Si shell burning.
This also reflects the decrease in the total Si mass in this region (see Figure \ref{fig:simass27LA}).
The energy release results in the expansion and the following contraction of above the Si-rich layer with $M_r \ga 2.1 M_\odot$.
This expansion and contraction can be also visible in the O mass fraction in the phase II (right panel of Figure \ref{fig:mfrav27LA_SiOinner}).

\begin{figure*}
\includegraphics[width=\textwidth]{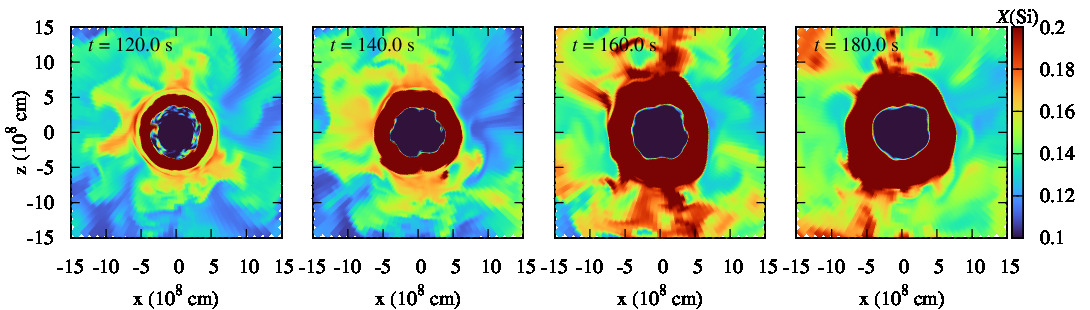}
\caption{Slices on the \textit{x-z} plane showing the Si mass 
fraction distribution of model 27\LA at 120, 140, 160, and 180 s from the left to the right.
The side length of the cubic box is $3.0 \times 10^{9}$ cm.
\label{fig:2Dmfsi}}
\end{figure*}

\begin{figure}
\vspace*{4mm}
\includegraphics[width=\columnwidth]{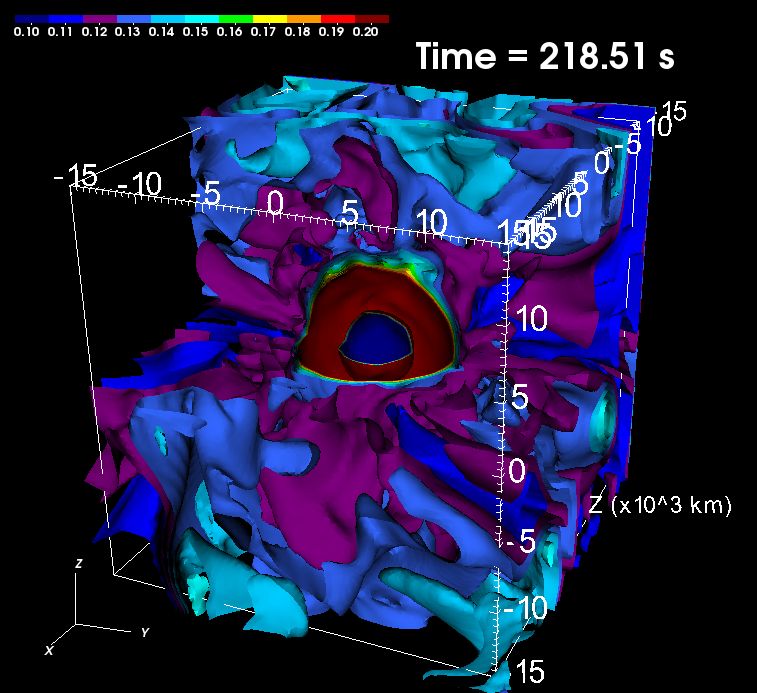}
\caption{The 3D contours the Si mass 
fraction distribution of model 27\LA at 218.51 s. 
The side length of the cubic box is $3.0 \times 10^{9}$ cm.
\label{fig:3Dmfsi}}
\end{figure}

\begin{figure*}
\includegraphics[width=\textwidth]{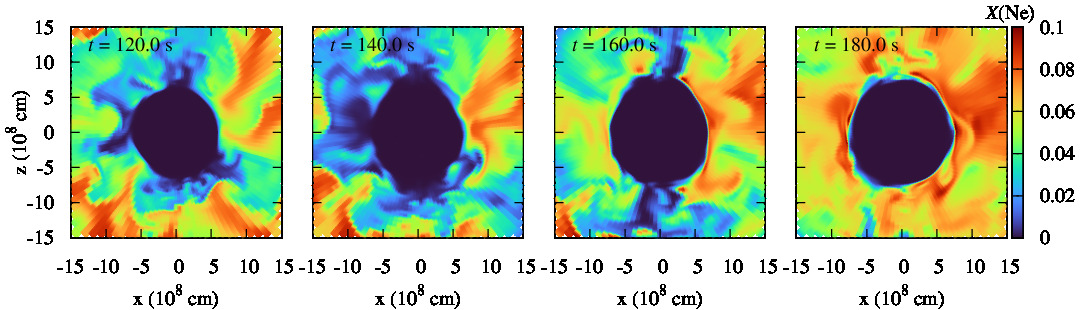}
\caption{Same as Figure \ref{fig:2Dmfsi}, but for the Ne mass fraction.
\label{fig:2Dmfne}}
\end{figure*}

\begin{figure}
\vspace*{2mm}
\includegraphics[width=\columnwidth]{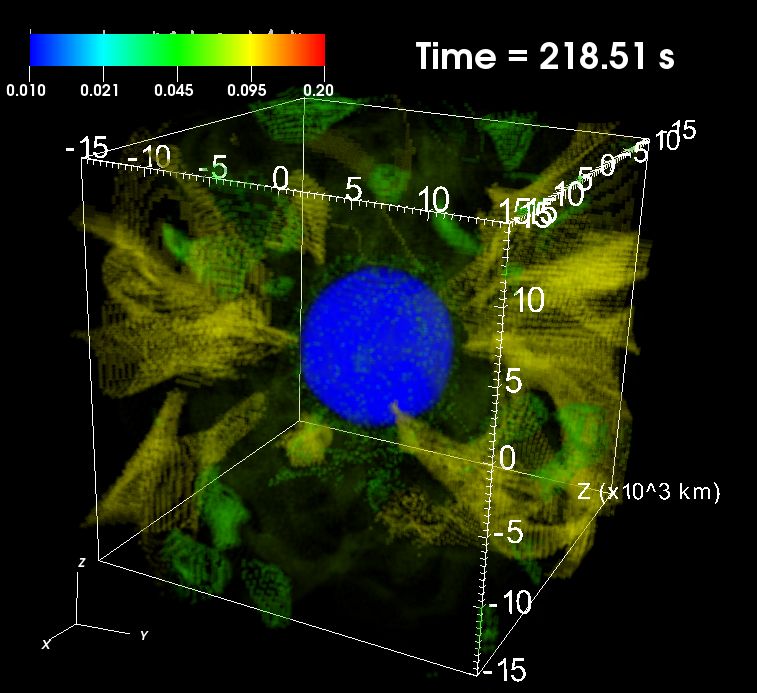}
\caption{Same as Figure \ref{fig:3Dmfsi}, but for the Ne mass fraction.  
\label{fig:3Dmfne}}
\end{figure}

In the phase III (127 s $\leq t \leq 218$ s), a dorm-like transient expansion and contraction can be seen in Figure \ref{fig:mfrav27LA_SiOinner}. 
This is predominantly triggered by O shell burning in the range of $M_r =$ 2--2.6 $M_{\odot}$, which corresponds to the bluish low O mass fraction regions in the phase III. 
Figure \ref{fig:omass27LA} shows the time evolution of the total O mass in the range of 2.0--2.6 $M_\odot$ where the time variation of the O mass fraction is mainly seen.
We also see the decrease in the oxygen mass in this region in 130--180 s.
On the other hand, the Si mass fraction increases in the region $M_r \sim 2.2 M_\odot$ by the O shell burning.
The Si mass in the region 1.8--2.2 $M_\odot$ also increases between $\sim$130--150 s (see Figure \ref{fig:simass27LA}).
Because of the O shell burning and inward turbulent mixing, the Si mass fraction increases in the dorm-like region concurrently (see the red region in the top right panel).

\paragraph{Silicon mass fraction in the O/Si/Ne layer}
Let us move to the viewpoint of the Si mass fraction in the O/Si/Ne layer.
The top left panel of Figure \ref{fig:mfrav27LA} is for the Si mass fraction for the entire O/Si/Ne layer.
Si is an ash of the Ne and O burning. 
By inspecting the Si mass fraction distribution, hereafter we try to understand the complicated interplay of the Ne burning (top right panel) and the O burning (middle right panel), which take place in a different manner.
 
In the phase I (0 s $\leq t \leq 107$ s), the increase of the Si mass fraction, which takes place from the near base region ($10 \times 10^{8}$ cm) to the outer region ($20 \times 10^{8}$ cm), is clearly visible.
This Si is produced through the Ne shell burning at the bottom of this layer.
By comparing with the extension of the turbulent activity shown by the bottom panels of Figure \ref{fig:Ma}, the correspondence between the two regions is confirmed.
This  naturally supports that the turbulent mixing globally takes place in the extended O/Ne/Si layer.

We note that the mass shells from 2 to 8 $M_\odot$ moves in a synchronized fashion in each phase.
This expansion and contraction propagates from the bottom of the Si/O-rich layer inward and outward with the sound velocity of (5--7)$\times 10^{8}$ and (2--5)$\times 10^{8}$ cm s$^{-1}$, respectively.
When an episodic burning occurs, the temperature and the pressure gradient at the burning region increase.
The regions inside and outside the burning shell expand to achieve a hydrostatic equilibrium.
In the region inside the Si/O-rich layer, the structural change propagates via the sound wave propagation within 1 s.

\paragraph{2D slices and 3D evolution}
Figure \ref{fig:2Dmfsi} depicts the evolution of the 2D slice on the \textit{x-z} plane for the Si mass fraction for four time snapshots from 120 s to 180 s in the phases II and III.
The 3D contours of the Si mass fraction at the last step are also shown in Figure \ref{fig:3Dmfsi}.
The size of the Si-rich region grows up through the O shell burning and the expansion as shown in Figure \ref{fig:mfrav27LA}.
In the phase III, the size of the inhomogeneities of the Si mass fraction in an inner region of the O/Si/Ne layer is shown to become bigger with time. This is because of the stronger turbulence mixing encompassing the whole layer (e.g., the bottom panels of Figure \ref{fig:Ma} for the region between the two horizontal white dashed lines in the phase III) predominantly triggered by the Ne shell burning (e.g., the bottom right panel of Figure \ref{fig:mfrav27LA}). 

Similar to Figure \ref{fig:2Dmfsi}, Figure \ref{fig:2Dmfne} shows 2D slices for the Ne mass fraction.
The 3D contours of the Ne mass fraction at the last step are also shown in Figure \ref{fig:3Dmfne}.
In the relatively early phase of the phase III (at $\sim$140 s), one can see yellow to red regions in the northeast and southwest directions, which shows the penetration of the high Ne fraction matter 
downward into the low Ne fraction region (greenish and bluish regions). 
This convection mixing develops more strongly with time (see the right panel).
In fact, the Ne mass fraction is  homogenized to a large extent in the O/Ne/Si layer as already shown in Figure \ref{fig:mfner27LA} (see the radial composition profile at $\ga$210 s).

\begin{figure}
\includegraphics[width=\columnwidth]{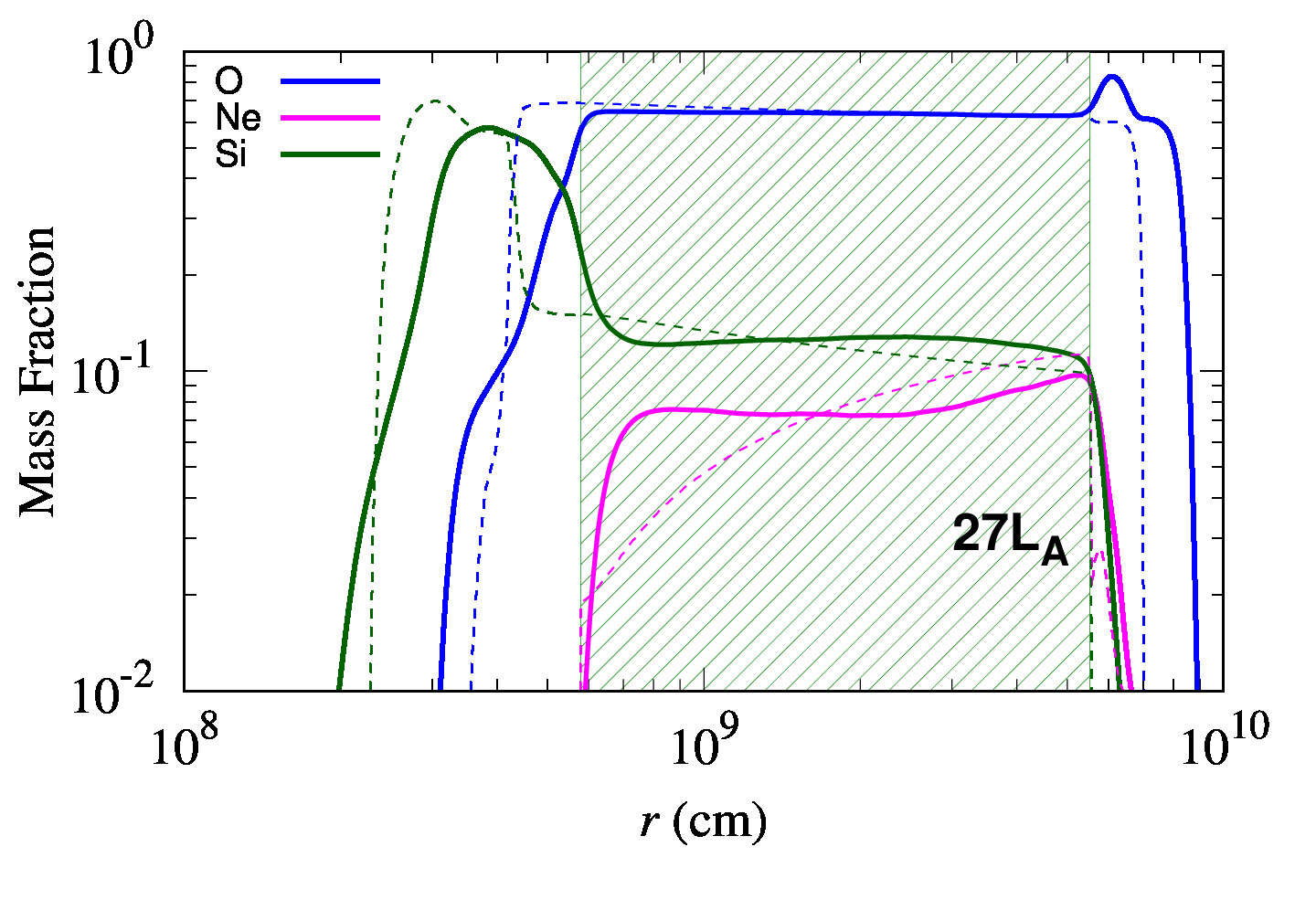}
\includegraphics[width=\columnwidth]{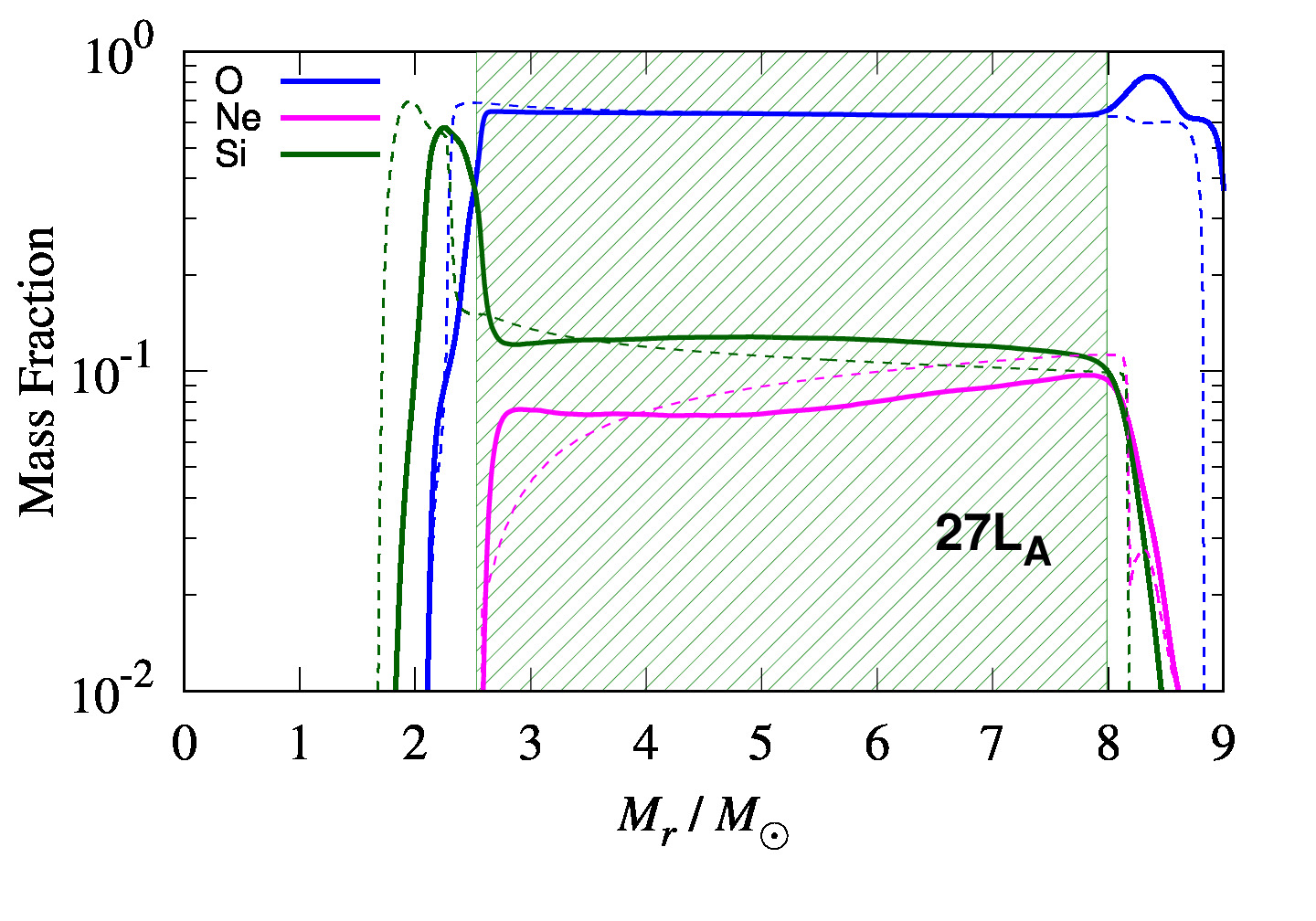}
\caption{Mass fraction profiles of the angle-averaged O, Ne, and Si at the initial (dashed lines) and last (solid lines) time steps of the 3D hydrodynamics simulations for model 27\LA.
Top and bottom panels correspond to the function of the radius and mass coordinate, respectively.
\label{fig:mfONeSi}}
\end{figure}

\paragraph{Radial composition profile and angular dispersion}
Finally, in Figure \ref{fig:mfONeSi},
we compare the composition distributions of representative $\alpha$-elements at the initial (dashed lines) and final time steps (solid lines) of the 3D simulations of model 27\LA. 
In the corresponding animations, we also present the accompanying two thin lines, which denote the angular dispersion relative to the angle-average one (thick line).
We note that in the 1D evolution the radial profiles of the O, Ne, and Si mass fractions scarcely change during the final $\sim$100 s up to the central temperature of $10^{10}$ K.

Regarding the O mass fraction (blue lines),  the angular dispersion is less than 2 \% in the O/Si/Ne layer (e.g., the shaded region of $7 \times 10^{8}$ cm $\lesssim r \lesssim 6 \times 10^{9}$ cm and 2.7 $M_\odot \lesssim M_r \lesssim 8.4 M_\odot$).
Since the O mass fraction is large and almost constant in this layer at the initial step of the simulation, the effect of turbulent mixing is not seen in the O mass fraction profile except for the region close to the base of the layer (as seen from the deviation from the thin dashed line and the thick solid line).

As is expected from the global mixing in the O/Si/Ne layer (Figures \ref{fig:2Dmfne} and \ref{fig:3Dmfne}), 
the initial Ne composition that gradually increases with the radius in the layer (see the dashed magenta line) is flattened 
owing to the mixing at the final simulation time (see the solid magenta line). 
This means that the 3D model shows smaller radial dependence than the 1D model for the composition profile 
in the O/Si/Ne layer at the last step.
Regarding the angular dispersion, it becomes larger compared to those of O (blue lines) and Si (green lines). 
It is about 10-20 \% in the inner layer of $r \sim$(7-20)$\times 10^8$ cm and $M_r \sim$ 2.7-4.4 $M_\odot$, reaching 20-30 \% near the upper boundary of the layer. 
Regarding the Si mass fraction (green lines), the initial distribution (thin dashed green line) is slightly shifted outward, 
and the initial composition gradient in the layer is also flattened owing to the mixing 
(compare the thin with thick green lines at $6\times 10^{8}$ cm $\lesssim r \lesssim 
6 \times 10^{9}$ cm and 2.6 $M_\odot \lesssim M_r \lesssim 8.4 M_\odot$). 
The angular dispersion is less than 10\%--20\% in the layer.

Irrespective of the composition differences, it can be commonly observed that the angular dispersion becomes bigger at the layer interface, where the nuclear burning occurs vigorously and the multidimensional effects become significant. In fact, the largest dispersion of 20\%--30\% of the Ne (magenta line) is obtained near the boundary between the O/Ne/Si layer ((6--7)$\times 10^{8}$ cm) and the O/C layer ($\sim$6$\times 10^{9}$ cm), and the dispersion of the O mass fraction distribution (blue line) becomes bigger at the interface between the O/Ne/Si layer and the Si layer inside.

The effect of shell burning on the turbulent mixing can be found from the time variation of the mass fraction profile in the animated version of Figure \ref{fig:mfONeSi}.
When a shell burning occurs, the radial distribution of angle-averaged mass fractions is homogenized.
On the other hand, large aspherical turbulence enhances the angular dispersion of mass fractions.
Then, the dispersion becomes smaller as the turbulence weakens.

We explain the variation of O and Ne mass fraction distributions in phase III as an example.
In 127 s $\le t \lesssim 150$ s, strong turbulence is activated in the Si/O layer with (4--6)$\times 10^{8}$ cm (see also the right panel of Figure \ref{fig:Ma}).
The steep rise of the O mass fraction with radius in this region becomes less steep, and at the same time, the O mass fraction decreases with time.
On the other hand, an increase in the angular dispersion is seen.
The turbulence weakens after $t \sim$150 s and this dispersion also becomes smaller (see the paragraph on Ne mass fraction).
For the Ne mass fraction, we see the radial gradient in the O/Si/Ne layer at the beginning of phase III.
The Ne mass fraction becomes gradually homogenized outward from $r \sim 7 \times 10^{8}$ cm, and the angular dispersion becomes larger.
The angular dispersion in the radially homogenized region is about 40\% at $t$ = 150 s.
This dispersion gradually decreases after this time.

In what follows,  we attempt to make an order-of-magnitude estimate to explore whether the 3D mixing (leading to the flattening of the initial composition distribution) could or could not be treated as a diffusion process, which an evolution code commonly does.
The initial 1D data have typical values of the convective velocity $v_{\rm cv} \sim 2 \times 10^{7}$ cm s$^{-1}$, the mixing length $l_{\rm cv} = \alpha H_p \sim 5.5 \times 10^{8}$ cm, and the diffusion coefficient $D_{\rm cv} = v_{\rm cv} l_{\rm cv}/3 \sim 3.6 \times 10^{15}$ cm$^2$ s$^{-1}$ in the O/Ne/Si layer of model 27\LA.
Thus, the diffusion timescale to homogenize the composition gradient is estimated as $\tau_{\rm cv} \sim L^2/D_{\rm cv} \sim 6.3 \times 10^2$ s, where the size of the composition gradient of Ne of $L \sim 1.5 \times 10^{9}$ cm is used.
This is much longer than the evolutionary timescale in this late phase, and it is why the original mass fraction distribution has a gradient within the convective region.
However, our 3D simulation results in the radially homogeneous distributions of the angle-averaged Ne mass fraction in a timescale of $\sim$100 s by the turbulent mixing.
We have noticed that this short timescale is rather compatible to the advection timescale of the turbulent flow, $\tau_{\rm turb} \sim L/v_{\rm turb} \sim 40$ s.
Much more systematic 3D stellar evolution modeling is preferable to seek for the method to mimic the multidimensional effects in mixing models in 1D stellar evolution models \citep[e.g.,][]{Couch15,bernhard16,bernhard18_prog,Yadav19}.

\begin{figure}
\includegraphics[width=\columnwidth]{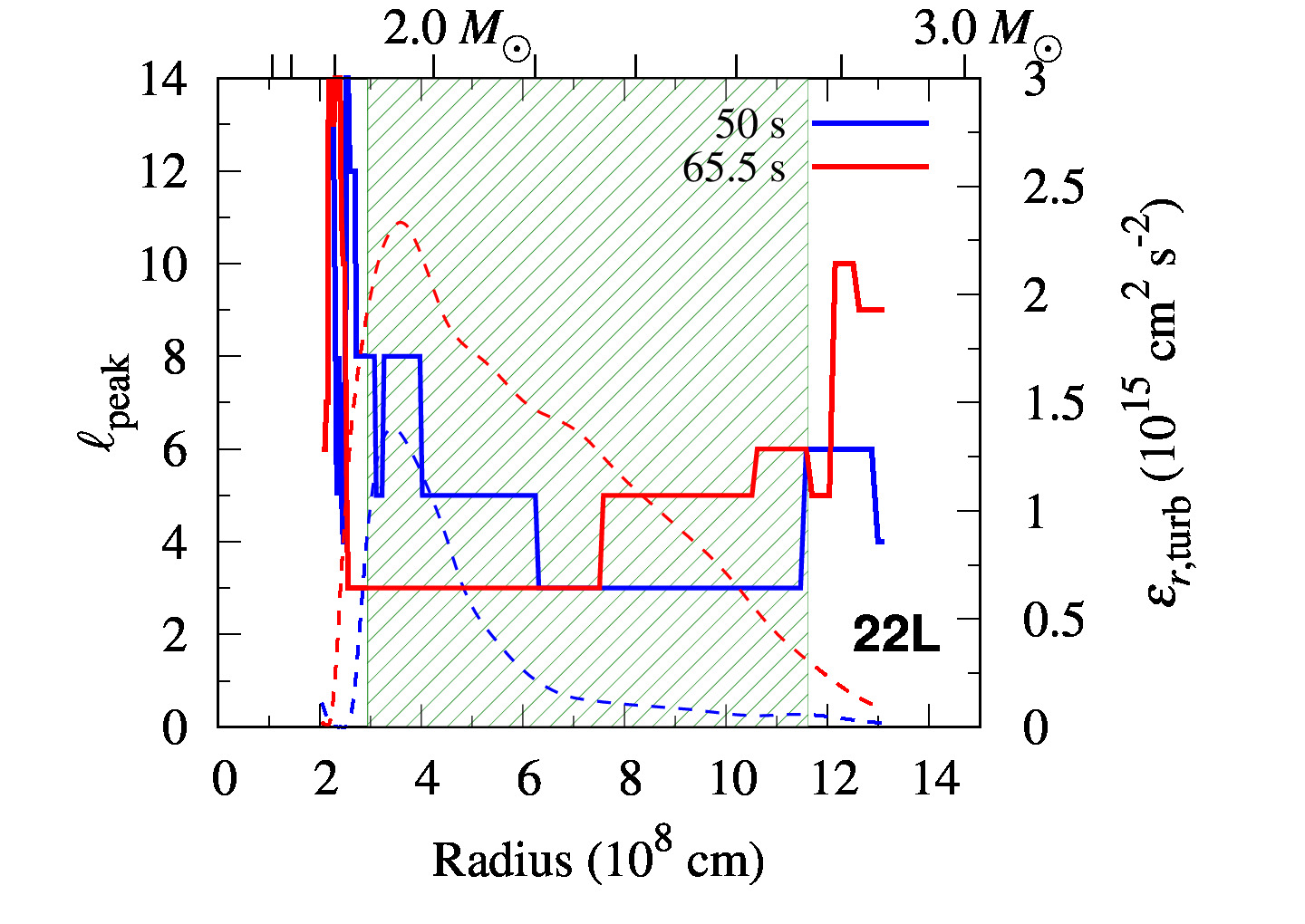}
\includegraphics[width=\columnwidth]{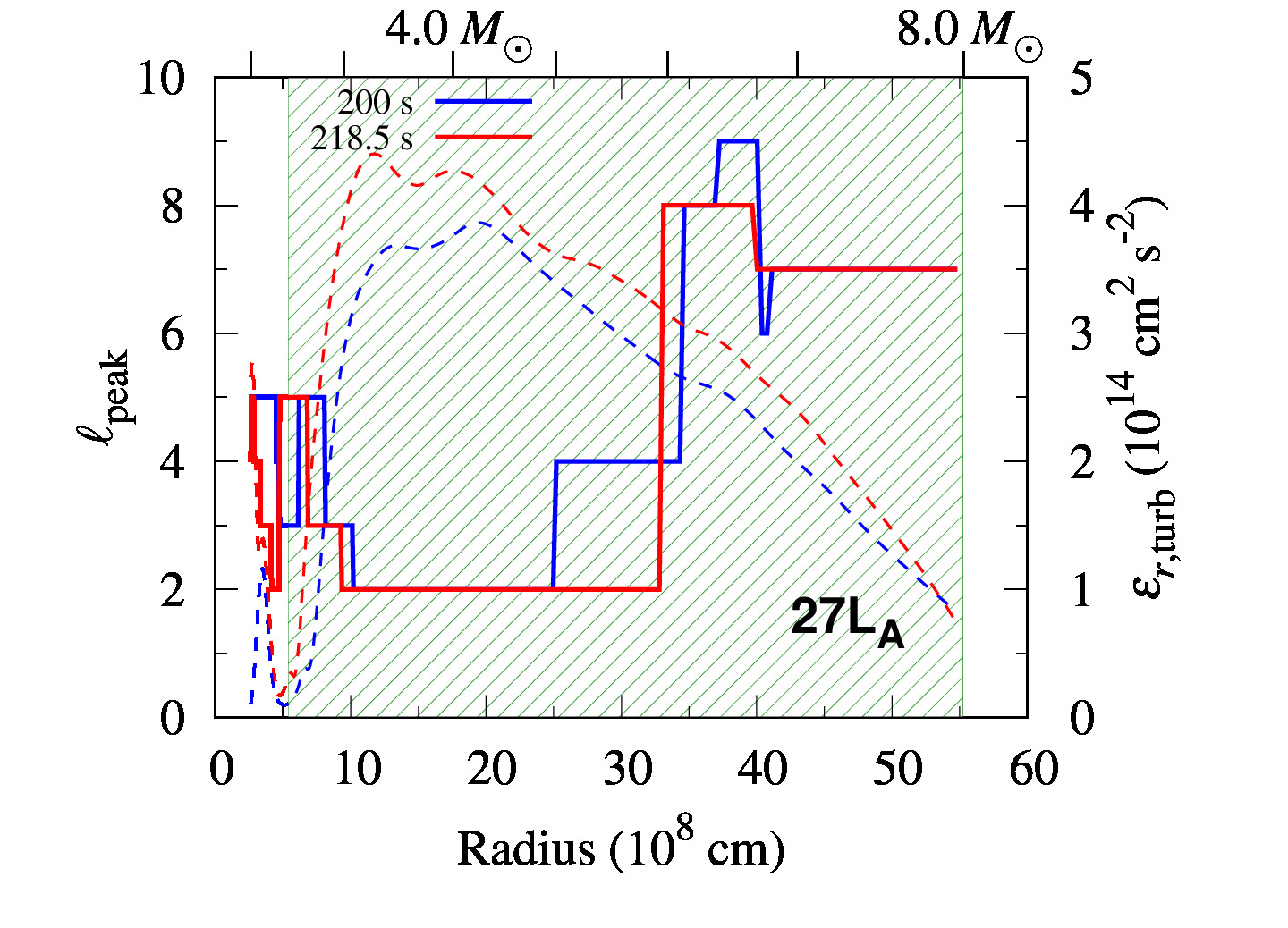}
\caption{Radial profiles of the peak mode for spherical harmonics $\ell_{\rm peak}(r)$ (solid lines with the $y$-axis on the left-hand side) and the turbulent energy $\epsilon_{r,{\rm turb}}(r)$ (dashed lines with the $y$-axis in the right-hand side) for models 22L (top panel) and 27L$_{\rm A}$ (bottom panel), respectively (see text for the definition).
The red curves indicate the quantities at the last step.
The blue curves indicate at 60 s for model 22L and 200 s for model 27\LA.
The green hatched region corresponds to that of
 Figure \ref{fig:mfONeSi}, showing the initial location of the Si/O and O/Si/Ne layers of models 22L and 27\LA.
\label{f9}}
\end{figure}

\subsection{Spectrum Analysis of Turbulence and the Spherical Harmonics Modes} \label{subsec:powerspectra}

In what follows, we propose a useful method
in order to extract the information about the typical eddy size of turbulence
 and its relation to the turbulent energy in the burning shells of our two 3D models. 
In our previous study \citep{Yoshida19}, we performed a spectrum analysis of the turbulent velocity at the radius where the maximum Mach number was obtained (e.g., Figure 6 in that paper). 
If we follow the conventional treatment in model 27\LA, the location of the maximum Mach number is identified at $\sim$3$\times 10^{9}$ cm (see the right panel of Figure \ref{fig:Ma}). 
On the other hand, the O/Si/Ne layer with the similar amplitude of the Mach number extends in a wider region up to $r \sim 6 \times 10^9$ cm near at the upper boundary of the layer (the upper horizontal white dashed lines). In such a widely extended shell, it should be better to study the radial dependence of the spectrum $c_{\ell}(r)$ of the turbulent velocity, instead of estimating the power spectrum at one radial location with the maximum Mach number.
  
We calculate the power spectrum of the radial turbulent velocity as 
\begin{equation}
   c_{\ell}^2(r) = \sum_{m = -\ell}^{\ell} \left| \int (v_r - \langle v_r \rangle)Y_{\ell m}^{*}(\theta,\phi)d\Omega \right|^2,
\end{equation}
where $Y_{\ell m}^{*}(\theta,\phi)$ is the
(complex conjugate) spherical harmonics of degree $\ell$ and order $m$ and $\ell$ is from 1 to 50.
We investigate the peak mode $\ell_{\rm peak}(r)$ as a function of radius.
The (specific) turbulent energy, the energy of the radial turbulent velocity, is evaluated as
\begin{equation}
    \varepsilon_{r,{\rm turb}}(r) = \frac{1}{8\pi} \sum_\ell c_\ell^2(r).
\end{equation}

Figure \ref{f9} shows the radial profiles of  $\ell_{\rm peak}(r)$ (solid lines) and the   turbulent energy $\varepsilon_{r,{\rm turb}}$ (dashed lines) at two representative times near (blue line) and at the final simulation time (red line) for models 22L (top panel) and 27\LA (bottom panel), respectively. 
From the top panel, one can see in model 22L that the value of $\ell_{\rm peak}(r)$ (solid lines) is in the range of $\sim$3--5 in most of the Si/O layer (shaded region). The smaller turbulent mode $\ell_{\rm peak}(r) \sim$ 3 is associated with the higher turbulent velocity (the dashed lines), whereas near at the outer boundary of the layer (e.g., the right edge of the shaded region) the turbulent mode becomes bigger ($\ell_{\rm peak}(r) = 7$) with the smaller turbulent energy (e.g., seen as a drop in the dashed lines). 
The high turbulent energy near at the base of the Si/O layer (see also the right panel of Figure \ref{fig:Ma}) is triggered by O shell burning. We consider that the low $\ell_{\rm peak}(r)$ is a natural outcome of the growing large-scale turbulence near at the base of the burning layer. 
One also sees the value of $\ell_{\rm peak} \sim 8$ at the bottom of the Si/O layer (see the blue solid line).
This corresponds to the start of the turbulence development by the ignition of the O shell burning.
As a side remark, the high $\ell_{\rm peak}(r)$ ($\lesssim$12) below the lower convective boundary may be not surprising because of the weak turbulent activity and also of the thinner width of the region.

\begin{deluxetable*}{lccccccccccc}[ht]
\tablecaption{Key Parameters for the SSH Decomposition of the Turbulent Mach Number}
\tablewidth{0pt}
\tablehead{
\colhead{Model} & \colhead{$r_{\rm in}$} & \colhead{$r_{\rm out}$} & \colhead{$\lambda_r$} & \colhead{$n_{\rm end}$} & \colhead{$\ell_{\rm end}$} & \colhead{$\langle {\rm \Mach}^2 \rangle^{1/2}_{\rm max}$} & \colhead{$r(\langle {\rm \Mach}^2 \rangle^{1/2}_{\rm max})$} & \colhead{$\ell_{\rm ave}$} &  \colhead{$\Delta L_{\rm crit}/L_{\rm crit}$} & \colhead{$r'_{\rm in}$} & \colhead{$r'_{\rm out}$} \\
\colhead{} & \colhead{(10$^{8}$ cm)} &  \colhead{(10$^{8}$ cm)} & \colhead{(10$^{8}$ cm)} & \colhead{} & \colhead{} & \colhead{} & \colhead{(10$^{8}$ cm)} & \colhead{} & \colhead{(\%)} & \colhead{(10$^{8}$ cm)} &  \colhead{(10$^{8}$ cm)}
}
\startdata
25M    & 2.30 & 12.99 & 0.5 & 42 & 48 & 0.172 & 5.48 & 4.99 & 8.1 & 2.49 & 11.40 \\
22L    & 2.13 & 12.99 & 0.5 & 43 & 47 & 0.158 & 3.93 & 5.08 & 7.3 & 2.14 & 11.64 \\
27\LA & 5.45 & 63.25 & 2.5 & 46 & 43 & 0.103 & 37.6 & 4.22 & 5.7 & 5.29 & 55.24 \\
\enddata
\tablecomments{
 $r_{\rm in}, r_{\rm out}, \lambda_r, n_{\rm end}, \ell_{\rm end}$; see text for definition, the maximum turbulent Mach number 
 ($\langle {\rm \Mach}^2 \rangle^{1/2}_{\rm max}$) and the corresponding radius ($r(\langle {\rm \Mach}^2 \rangle^{1/2}_{\rm max})$), 
 the averaged spectrum number ($\ell_{\rm ave})$, the estimated reduction percentile of the critical luminosity 
 ($\Delta L_{\rm crit}/L_{\rm crit}$)  required  for an onset of neutrino-powered explosions (see text for more details), 
 and the radii of the inner ($r'_{\rm in}$) and outer ($r'_{\rm out}$) boundaries of the Si/O-rich layer of 1D stellar evolution models 
 at the last step.}
\label{tab:harm_param}
\end{deluxetable*}

The above trend is similar also to model 27\LA (bottom panel of Figure \ref{f9}). 
The minimum $\ell_{\rm peak}(r) = 2$ is obtained near at the base of the O/Si/Ne layer, which is smaller than model 22L, whereas $\ell_{\rm peak}(r)$ is found to be slightly higher ($\sim$8) near at the outer boundary than for model 22L ($\ell_{\rm peak}(r) \sim 5$).

In order to characterize a typical scale of  turbulent flows in the whole Si/O or O/Si/Ne layer (for models 22L and 27\LA, respectively),  we define the average mode of $\ell_{\rm peak}$, which we obtain from $\ell_{\rm peak}(r)$ by integrating with a weight of the turbulent energy as follows:
\begin{equation}
    \langle \ell_{\rm peak} \rangle = \frac{\int \ell_{\rm peak}(r) \rho(r) \varepsilon_{r,{\rm turb}}(r) r^2dr}{\int  \rho(r) \varepsilon_{r,{\rm turb}}(r) r^2dr},
    \label{lave}
\end{equation}
where $\rho(r)$ is the angle-averaged density as a function of radius.
The obtained $\langle \ell_{\rm peak} \rangle$ values at the last simulation time are 3.75 and 3.56 for models 22L and 27\LA, respectively. Using these number,  we touch on in the next section how much the difference in the turbulent modes ($\ell_{\rm peak}$), as well as the associated Mach number, could possible affect the explodability of these progenitors.

\subsection{Scalar Spherical Harmonics Decomposition of Turbulent Mach Number } \label{subsec:harmonics}

\begin{figure}
\includegraphics[width=\columnwidth]{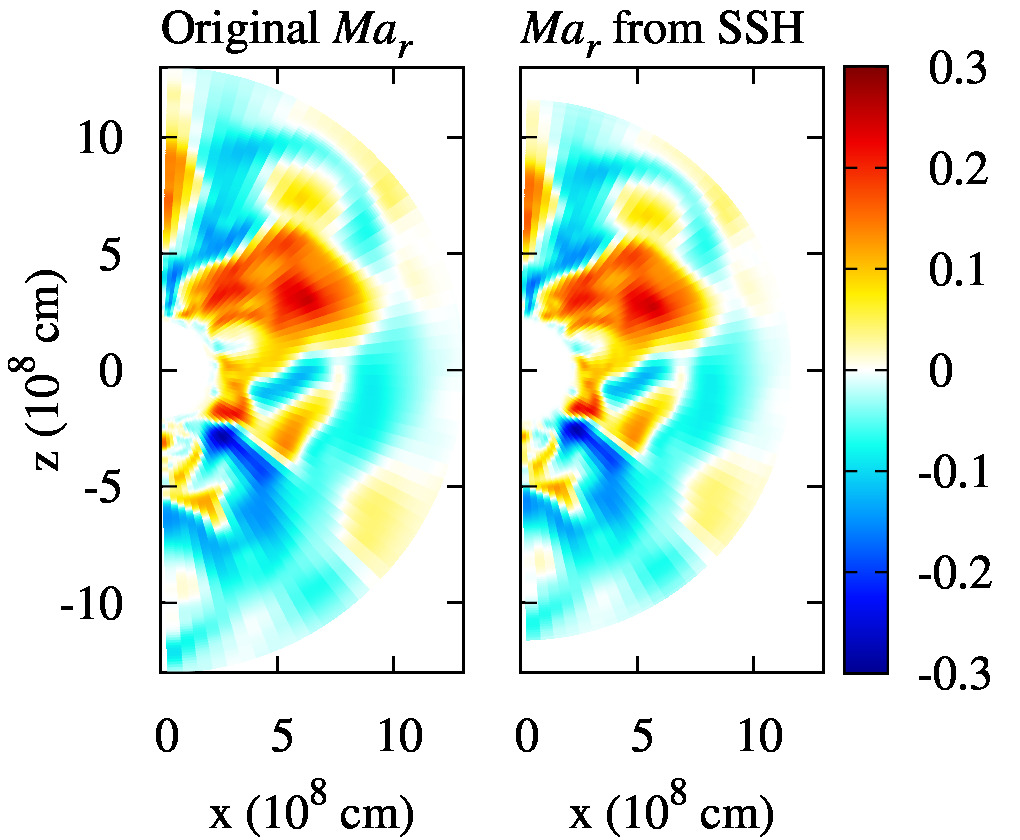}
\includegraphics[width=\columnwidth]{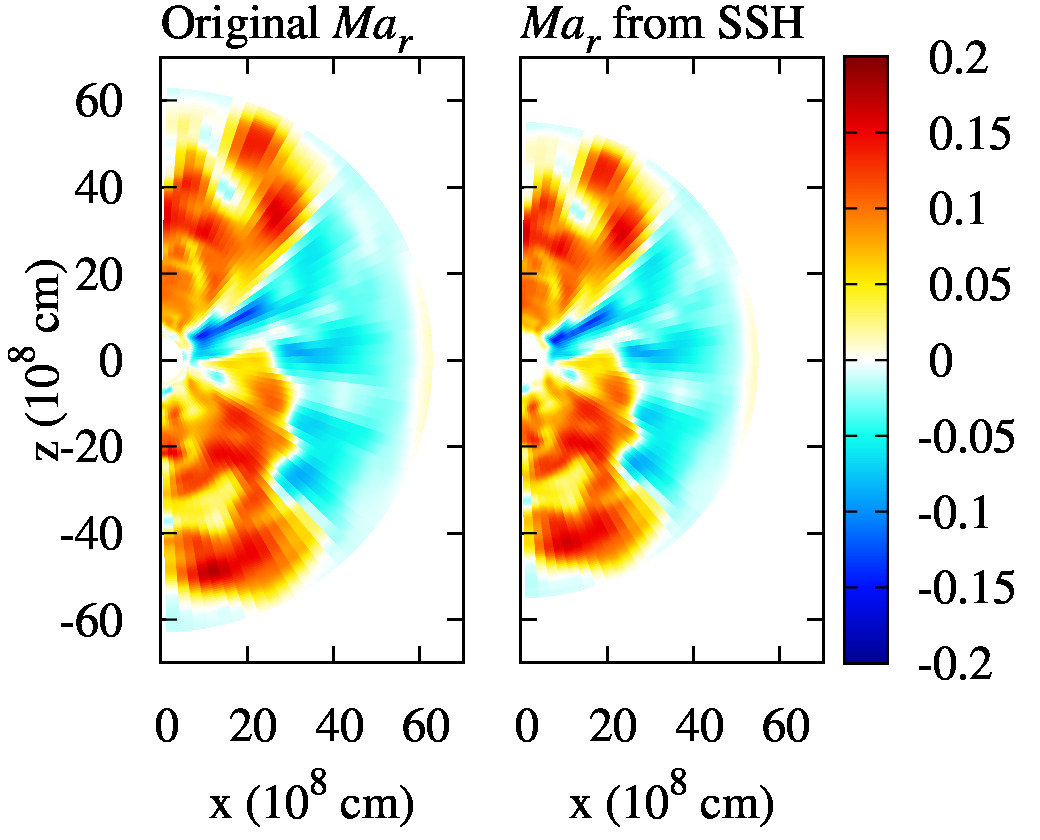}
\caption{Distributions of radial Mach number $\Mach_r$ at $\phi = 0$ in models 22L (top panels) and 27\LA (bottom panels).
The left panel is the original $\Mach_r$ distribution obtained from our 3D model, whereas the right panel is the $\Mach_r$ distribution reconstructed from the result of SSH analysis and fitted to the corresponding range at the last step of the 1D model.
\label{f10}}
\end{figure}

Bearing in mind an implementation of the  precollapse inhomogeneities (obtained in this work) in future CC SN simulations, we present an analysis of the SSH mode of turbulent Mach number. Following \citet{chatz14}, we evaluate the SSH decomposition of the radial Mach number ($\Mach_{r}$) distribution in the burning shells of models 22L and 27L$_{\rm A}$ as well as model 25M in our previous study \citep{Yoshida19}.

Following the procedure in the Appendix of \citet{chatz14}, one first needs to determine the inner ($r_{\rm in}$) and outer ($r_{\rm out}$) boundaries of the region of interest.
To determine the inner and outer boundaries, we first determine the radius of the maximum turbulent Mach number  $r_{{\rm Ma,max}}$ in the Si/O-rich layer.
Then, the inner boundary is determined as the location where the turbulent Mach number becomes a minimum when we go inward from $r_{{\rm Ma,max}}$.
The outer boundary is determined as the location where the turbulent Mach number becomes less than $1/3 \langle \Mach^2 \rangle^{1/2}_{\rm max}$ when we go outward from $r_{{\rm Ma,max}}$.
There is no local minimum around the outer boundary of the Si/O-rich layer.
The Mach number is $\la 0.1 \langle \Mach^2 \rangle^{1/2}_{\rm max}$ even outside the Si/O-rich layer.
Hence, we set the location of the outer boundary using the ratio to the maximum Mach number.
The radii of the boundaries are listed in Table \ref{tab:harm_param}.
We also set the shortest physical length scale $\lambda_r$ for these models in Table \ref{tab:harm_param}.
We set this scale to obtain the reconstruction factor, described later, below 0.1.
With the given boundaries and shortest physical length, the maximum numbers of the $n$-modes, $n_{\rm end}$, and $\ell$-modes, $\ell_{\rm end}$ in the SSH power spectra are determined as
\begin{equation}
    n_{\rm end} = \frac{2 (r_{\rm out} - r_{\rm in})}{\lambda_r},
    \label{eq:nend}
\end{equation}
and
\begin{equation}
    \ell_{\rm end} = \frac{\pi (r_{\rm out} + r_{\rm in})}{2 \lambda_r}.
    \label{eq:lend}
\end{equation}
The values of $n_{\rm end}$ and $\ell_{\rm end}$  are also listed in Table \ref{tab:harm_param}.
Then, we evaluate the set of the components of the SSH decomposition $a_{n\ell m}$ with a mode $n$, $\ell$, $m$ from the radial Mach number distribution Ma$_r(r,\theta,\phi)$ in the range between $r_{\rm in}$ and $r_{\rm out}$.
Details for the SSH decomposition are written in Appendix A.
The SSH decomposition components $a_{n\ell m}$ of models 25M, 22L, and 27\LA are listed in 
Tables \ref{tab:anlm_25M}--\ref{tab:anlm_27LA}, respectively, in Appendix A.
The machine-readable tables are also available.

When one applies the spatial distribution of the radial Mach number to a 1D stellar evolution model having an Si/O-rich layer in the range between $r'_{\rm in}$ and $r'_{\rm out}$, one can calculate it using the modified SSH decomposition components $a'_{n\ell m}$ (see Eq. (\ref{eq:mod-components}) in Appendix A).
In Figure \ref{f10}, we compare the original radial Mach number distribution (left panel) with the one that is reconstructed by the SSH method (right panel) for models 22L and 27\LA.
Note that the radial range of the Si/O-rich layer of the reconstructed one is fitted to the 1D model, which is listed in Table \ref{tab:harm_param}.
As seen, we obtain a nice match of a global feature. 
To quantify the difference, we estimate the reconstruction factor, which we estimate in the same range as
\begin{equation}
    f_{\rm rec} = \sqrt{
    \frac{\int (\Mach_r'({\bf x})-\Mach_r({\bf x}))^2 dV}
    {\int \Mach_r({\bf x})^2 dV}},
    \label{rec}
\end{equation}
where $\Mach_r(\bf x)$ and $\Mach_r'({\bf x})$ are the original and reconstructed radial Mach number, respectively (e.g.,  \citet{chatz14}).
The values of the reconstruction factor are 0.049, 0.056, and 0.065 for models 25M, 22L, and 27\LA, respectively.
Note that in the volume integration in Equation (\ref{rec}), we do not include the regions of five meshes from the inner and outer boundaries.
This is because the original Mach number has nonzero values at the boundary, whereas the reconstructed Mach number is set to be zero there by the imposed boundary conditions.
Some differences between the reconstructed Mach number and the original one with $\ga$0.1 appear in a small region close to the boundary of along the z-axis.
Given this, we consider that the 5\%--7\% level of the mismatch is not a big concern. 

Finally, we give an exploratory discussion of how the inhomogeneities and turbulence obtained in this work may impact the perturbation-aided explosion. 
For this purpose, we estimate the reduction rate of the critical neutrino luminosity required for the onset of neutrino-driven explosion (Equation of (9) in \citet{Collins18}), namely, $\Delta L_{\rm crit}/L_{\rm crit} \sim 2.34\, {\rm \Mach}/{\ell}$, with $\Mach$ and $\ell$ denoting the Mach number and the mode in the burning shells, respectively. 
Here we take $\ell$ from $\ell_{\rm ave}$ of Equation (\ref{lave}) and $\Mach$ from the maximum Mach number obtained in the computational domain (see Table \ref{tab:harm_param}). From this optimistic and crude estimate, we can expect $\sim$6\%--8\% levels of reduction in the critical neutrino luminosity. 
As already discussed in \citet{bernhard16_prog} and \citet{Collins18}, the expected reduction of $\sim$5\% could still make the turbulent perturbation in the burning shells one of several key ingredients for robust explosions. To put the final word, one needs to perform 3D SN simulations using 3D progenitors \citep{couch_ott15,bernhard18_prog,Bollig20}, which we leave for future study.

\section{Summary and Discussions} \label{sec4}

We performed 3D hydrodynamic simulations of the Si/O-rich layer just before the CC of two evolved massive stars having a wide Si/O-rich layer.
Model 22L has an Si/O layer with a width of $\sim$8$\times 10^{8}$ cm.
Model 27\LA has an O/Si/Ne layer with a width of $\sim$5$\times 10^{9}$ cm, which has been evolved from the O/Ne layer.
Although the width of the Si/O-rich layer is quite different between the two models, the turbulent Mach numbers reach $\langle \Mach^2 \rangle^{1/2} \ga 0.1$ owing to oxygen and neon shell burning.
In model 22L, turbulence is developed in the whole Si/O layer in a few tens of seconds, similarly to model 25M.
The angle-averaged Mach number rises by oxygen shell burning and finally reaches $\sim$0.15.
In model 27\LA, the convective region with $\langle \Mach^2 \rangle^{1/2} \sim 0.07$ 
extends outward from the bottom of the O/Si/Ne layer with time.
The neon shell burning that occurred in $\sim$130 s enhances the turbulence with $\langle \Mach^2 \rangle^{1/2} \sim 0.1$.
The enhanced turbulence is extended to almost all regions of the O/Si/Ne layer.
The spectrum expansion of the radial turbulent velocity results in the peak mode with $\ell_{\rm peak} \sim$ 2--3.
As shown in \citet{bernhard16_prog}, a low-mode-dominated turbulent velocity distribution will enhance the explodability of CC SNe.

The time evolution of the composition distribution is somewhat different between the two models owing to their different initial composition distributions.
In model 22L, the O mass fraction decreases from the bottom of the Si/O layer through the oxygen shell burning.
Although the turbulent mixing tends to homogenize the composition distribution, a small radial gradient remains. 
In model 27\LA, the turbulent motion induced by neon shell burning from $\sim$130 s tends to homogenize the chemical composition in the wide range of the O/Si/Ne layer.
We see the enhancement of the Si mass fraction up to $\sim$3$\times 10^9$ cm and inflow of the Ne rich materials to the bottom of the O/Si layer.
Such turbulent motions homogenize the radial mass fraction distribution.
In order to mimic the chemically homogeneous distribution obtained in the 3D simulations, a larger value of the diffusion coefficient would be necessary.
On the other hand, deviations of 10\%--30\% in the angular direction are formed in the O/Si/Ne layer.

Spatial distribution of the dominant mode obtained by power spectral analysis of the radial turbulent velocity shows a different feature between the two models.
We obtained a small radial dependence of the dominant mode throughout the Si/O layer in model 22L.
On the other hand, a radial dependence is shown in model 27\LA.
Namely, while the dominant mode is roughly constant inside $2 \times 10^9$ cm, where the turbulent velocity is higher, the turbulent velocity decreases and the dominant mode increases with increasing radius in the outer region.
When we consider the radial averaged power weighted by the radial turbulent velocity, the powers are 5.08 and 4.22 for models 22L and 27\LA, respectively.
Hence, a typical mode of the turbulent motion is not so different between the two models.
Taking account of the radial distribution of the radial turbulent velocity, we consider that a typical eddy size of the turbulent motion is several $\times 10^8$ to $\sim$10$^9$ cm and that such a scale of eddies would provide a favorable condition to the onset of neutrino-driven explosions.

In the current 3D simulations of stellar evolution, we still have a difficulty in the initial structure of 3D hydrodynamics simulations that a hydrostatic equilibrium is not established initially when we map the 1D stellar structure into a 3D simulation.
This would affect the difference of collapsing time between 1D and 3D simulations.
One of the reasons is that the number of adopted nuclear species is different between 1D (300) and 3D (21) stellar evolution simulations.
This problem will be improved by using a nuclear reaction network common to 1D and 3D simulations, which we leave for future work.

In this paper, we have calculated the evolution for 65.5 and 218.5 s for models 22L and 27\LA, respectively, until the central temperature becomes 10$^{10}$ K. Needless to say, a more long-term 3D simulation \citep[e.g.,][]{Yadav19}, albeit computationally expensive, should be done to fully capture the development of turbulent motion in the convective Si/O-rich layer.
In addition, the current angular resolution of our 3D models is $\delta \theta \sim 3^\circ$.
This limited resolution would suppress the development of the low $\ell$ mode of the turbulent velocity \citep{bernhard16_prog}.
These shortcomings should be also improved in our future work.

We have presented an analysis of SSH to add 3D characteristics of the velocity field to 1D pre-SN structure.
Currently, this analysis is performed only for the radial velocity field.
Asphericity of other quantities such as entropy field and chemical composition would also affect the explodability of SNe.
The reconstructed 3D velocity field is not a solution of the 3D hydrodynamic equation.
Adopting the obtained 3D structure to the initial structure of an SN model is the best way when one performs a 3D SN simulation.
However, when we adopt a 3D feature to a 1D structure data, this SSH analysis would be more realistic than introducing a parameterized asymmetry structure \citep[e.g.,][]{bernhard15}.
In the future, we plan to update the analysis to deal with vector spherical harmonics in 3D, which has been first investigated in 2D by \citet{chatz16}.

\acknowledgments

We thank the anonymous referee for reading carefully our manuscript and giving us many valuable comments and suggestions to improve this manuscript.
This study was supported in part by the Grants-in-Aid for the Scientific Research of the Japan Society for the Promotion of Science (JSPS) KAKENHI grant Nos. JP17H05206, JP17K14306, and
JP17H01130, JP17H06364, JP18H01212 (K.K. and T.T.) and 20H05249,
by the Central Research Institute of Explosive Stellar Phenomena (REISEP) of Fukuoka University and the associated project (No. 207002), and by JICFuS as a priority issue to be tackled by using the Post `K' Computer.
The numerical simulations were done using XC50 at the Center for Computational Astrophysics at the National Astronomical Observatory of Japan.

\software{HOSHI \citep{Takahashi16,Takahashi18,Takahashi19,Yoshida19}, 3DnSNe \citep{Nakamura16,Takiwaki16,Kotake18},
subroutines {\tt plgndr\_func} (modified from {\tt plgndr}), {\tt sphbes, bessjy, beschb} and function {\tt chebev} \citep{Press92}.}

\appendix

\begin{deluxetable*}{rrccccccc}[th]
\vspace*{-1cm}
\tablecaption{The SSH Decomposition Components $a_{n\ell m}$ for Model 25M \label{tab:anlm_25M}}
\tablewidth{0pt}
\tablehead{
\colhead{$\ell$} & \colhead{$m$} & \colhead{R or I} & \colhead{$a_{0\ell m}$} & \colhead{$a_{1\ell m}$} & \colhead{$a_{2\ell m}$} & \colhead{$a_{3\ell m}$} & \colhead{$a_{4\ell m}$} & \colhead{$a_{5\ell m}$}
}
\startdata
0 & 0 & R &  4.10E$-$02 &  1.78E$-$02 & $-$1.06E$-$02 &  1.54E$-$02 & $-$7.81E$-$03 &  8.27E$-$03 \\
0 & 0 & I &  0.00E+00 &  0.00E+00 &  0.00E+00 &  0.00E+00 &  0.00E+00 &  0.00E+00 \\
1 & 0 & R &  7.12E$-$02 &  1.01E$-$02 &  3.07E$-$02 &  1.08E$-$02 & $-$2.93E$-$03 & $-$1.56E$-$02 \\
1 & 0 & I &  0.00E+00 &  0.00E+00 &  0.00E+00 &  0.00E+00 &  0.00E+00 &  0.00E+00 \\
1 & 1 & R &  5.72E$-$01 & $-$2.09E$-$01 &  6.14E$-$02 & $-$7.90E$-$02 &  6.18E$-$02 & $-$4.44E$-$02 \\
1 & 1 & I &  1.32E$-$01 & $-$1.48E$-$02 &  1.45E$-$02 &  4.74E$-$02 &  6.34E$-$03 &  2.16E$-$03 \\
\enddata
\tablecomments{
In the third column, ``R" and ``I" mean real and imaginary part of $a_{n\ell m}$.
There are 46 columns in the entire table.
$(n_{\rm end}, \ell_{\rm end})$ = (42, 48).
The range of $m$ in this table is between 0 and $\ell$ for each $\ell$. \\
(This Table is available in its entirety in machine-readable form.)
}
\end{deluxetable*}

\begin{deluxetable*}{rrccccccc}[th]
\tablecaption{The SSH Decomposition Components $a_{n\ell m}$ for Model 22L
\label{tab:anlm_22L}}
\tablewidth{0pt}
\tablehead{
\colhead{$\ell$} & \colhead{$m$} & \colhead{R or I} & \colhead{$a_{0\ell m}$} & \colhead{$a_{1\ell m}$} & \colhead{$a_{2\ell m}$} & \colhead{$a_{3\ell m}$} & \colhead{$a_{4\ell m}$} & \colhead{$a_{5\ell m}$}
}
\startdata
0 & 0 & R &  2.45E$-$02 &  1.16E$-$02 & $-$9.50E$-$03 &  1.44E$-$02 & $-$7.08E$-$03 &  6.23E$-$03 \\
0 & 0 & I &  0.00E+00 &  0.00E+00 &  0.00E+00 &  0.00E+00 &  0.00E+00 &  0.00E+00 \\
1 & 0 & R & $-$2.48E$-$01 & $-$1.23E$-$02 &  6.53E$-$02 &  2.18E$-$02 &  9.52E$-$03 &  3.04E$-$02 \\
1 & 0 & I &  0.00E+00 &  0.00E+00 &  0.00E+00 &  0.00E+00 &  0.00E+00 &  0.00E+00 \\
1 & 1 & R & $-$5.79E$-$02 &  1.32E$-$01 &  3.08E$-$02 &  6.15E$-$02 & $-$1.61E$-$02 &  2.15E$-$02 \\
1 & 1 & I &  6.04E$-$02 &  1.55E$-$02 &  8.26E$-$03 & $-$1.10E$-$02 &  5.05E$-$03 & $-$1.82E$-$02 \\
\enddata
\tablecomments{
In the third column, ``R" and ``I" mean real and imaginary part of $a_{n\ell m}$.
There are 47 columns in the entire table.
$(n_{\rm end}, \ell_{\rm end})$ = (43, 47).
The range of $m$ in this table is between 0 and $\ell$ for each $\ell$. \\
(This table is available in its entirety in machine-readable form.)
}
\end{deluxetable*}

\begin{deluxetable*}{rrccccccc}[h]
\vspace*{-1cm}
\tablecaption{The SSH Decomposition Components $a_{n\ell m}$ for Model 27\LA \label{tab:anlm_27LA}}
\tablewidth{0pt}
\tablehead{
\colhead{$\ell$} & \colhead{$m$} & \colhead{R or I} & \colhead{$a_{0\ell m}$} & \colhead{$a_{1\ell m}$} & \colhead{$a_{2\ell m}$} & \colhead{$a_{3\ell m}$} & \colhead{$a_{4\ell m}$} & \colhead{$a_{5\ell m}$}
}
\startdata
0 & 0 & R &  1.51E$-$01 & $-$6.12E$-$02 & $-$1.56E$-$02 &  2.24E$-$02 & $-$3.50E$-$02 &  3.86E$-$02 \\
0 & 0 & I &  0.00E+00 &  0.00E+00 &  0.00E+00 &  0.00E+00 &  0.00E+00 &  0.00E+00 \\
1 & 0 & R & $-$2.70E+00 &  1.21E$-$01 & $-$3.93E$-$01 &  9.22E$-$01 & $-$1.38E$-$01 &  2.92E$-$01 \\
1 & 0 & I &  0.00E+00 &  0.00E+00 &  0.00E+00 &  0.00E+00 &  0.00E+00 &  0.00E+00 \\
1 & 1 & R &  4.39E$-$02 &  5.77E$-$01 & $-$4.68E$-$02 & $-$4.02E$-$02 & $-$2.46E$-$01 &  1.78E$-$01 \\
1 & 1 & I & $-$7.59E$-$01 &  2.67E$-$01 & $-$3.28E$-$01 &  2.01E$-$01 & $-$8.54E$-$02 & $-$2.93E$-$03 \\
\enddata
\tablecomments{
In the third column, ``R" and ``I" mean real and imaginary part of $a_{n\ell m}$.
There are 50 columns in the entire table.
$(n_{\rm end}, \ell_{\rm end})$ = (46, 43).
The range of $m$ in this table is between 0 and $\ell$ for each $\ell$. \\
(This table is available in its entirety in machine-readable form.)
}
\end{deluxetable*}

\section{Scalar Spherical Harmonics Decomposition}

We describe the SSH decomposition of the radial Mach number distribution in a convective region.
We assume here that the turbulent Mach number is zero at the radii of the inner and outer boundaries in the convective region.
Hence, we adopt the general solution of the Helmholtz equation with Dirichlet boundary conditions in the spherical coordinates.
The turbulent radial Mach number at the coordinate $\Mach(r, \theta, \phi)$ is written as a superposition of eigenfunctions
\begin{equation}
    \Mach_r(r,\theta,\phi) = \sum_{\ell=0}^{\ell_{\rm end}} \sum_{m=-\ell}^{\ell} \sum_{n=0}^{n_{\rm end}} a_{n\ell m} g_{\ell}(k_{n\ell}r) Y_{\ell m}(\theta,\phi),
    \label{aeq:decomposition}
\end{equation}
where $Y_{\ell m}(\theta,\phi)$ is a spherical harmonics function and $g_\ell(k_{n\ell}r)$ is a solution of the separated radial component of the Helmholtz equation,
\begin{equation}
    \frac{1}{r^2}\frac{d}{dr}\left(r^2 \frac{dg_{\ell}(k_{n\ell}r)}{dr}\right) +
    \left(k_{n\ell}^2 - \frac{\ell (\ell +1)}{r^2}\right) g_{\ell}(k_{n\ell}r) = 0.
    \label{aeq:radial_equation}
\end{equation}
The numbers $n_{\rm end}$ and $\ell_{\rm end}$ are determined from Eqs. (\ref{eq:nend}) and (\ref{eq:lend}) in the main text. 
The Dirichlet boundary conditions at the radii of the inner boundary  $r_{\rm in}$ and the outer boundary $r_{\rm out}$ are
\begin{equation}
    g_{\ell}(k_{n\ell}r_{\rm in}) = g_{\ell}(k_{n\ell} r_{\rm out}) = 0.
    \label{aeq:radial_boundary}
\end{equation}
The ($n,\ell$)-component of the radial eigenfunction can be written using the first and second Bessel functions $j_{\ell}(k_{n\ell}r)$ and $y_{\ell}(k_{n\ell}r)$ as
\begin{equation}
    g_{\ell}(k_{n\ell}r) = N_{n\ell} \left\{ y_{\ell}(k_{n\ell}r_{\rm in}) j_{\ell}(k_{n\ell}r) - j_{\ell}(k_{n\ell}r_{\rm in}) y_{\ell}(k_{n\ell}r) \right\},
    \label{aeq:g_nlkr}
\end{equation}
where $N_{n\ell}$ is the normalization constant.
This form of the function satisfies the inner boundary condition.
The wavenumber $k_{n\ell}$ is determined to satisfy the outer boundary condition.
The normalization constant is obtained from the orthogonality condition of
\begin{equation}
    \int_{r_{\rm in}}^{r_{\rm out}} g_{\ell}(k_{n\ell}r) g_{\ell}(k_{m\ell}r) r^2 dr = \delta_{nm},
    \label{aeq:radial_orthol}
\end{equation}
where $\delta_{nm}$ is the Kronecker symbol.
The normalization constant is calculated as
\begin{eqnarray}
    N_{n\ell} &=& \left[\frac{r_{\rm out}^3}{2} \left\{  y_{\ell}(k_{n\ell}r_{\rm in}) j_{\ell-1}(k_{n\ell-1 }r_{\rm out}) \right. \right. \nonumber \\
     &&  -  j_{\ell}(k_{n\ell}r_{\rm in}) y_{\ell-1}(k_{n\ell-1}r_{\rm out}) \}^2  \nonumber \\
    && - \left. \frac{r_{\rm in}^3}{2} \left\{ y_{\ell}(k_{n\ell}r_{\rm in}) j_{\ell-1}(k_{n\ell-1}r_{\rm in}) \right. \right. \nonumber \\
    && -  \left. \left. j_{\ell}(k_{n\ell}r_{\rm in}) y_{\ell-1}(k_{n\ell-1}r_{\rm in}) \right\}^2 \right]^{-1/2}. 
     \label{aeq:radial_normalization}
\end{eqnarray}
The decomposed components of the radial Mach number $\Mach_r(r,\theta,\phi)$ by SSH $a_{n\ell m}$ are calculated as
\begin{equation}
    a_{n\ell m} = \int_{r_{\rm in}}^{r_{\rm out}} \Mach_r(r,\theta,\phi) g_{\ell}(k_{n\ell}r) Y^{\ast}_{\ell m}(\theta,\phi) r^2 dr d\Omega,
    \label{aeq:components}
\end{equation}
where the asterisk denotes complex conjugation.

As for the information of aspherical distributions, we prepared the SSH decomposition components of the radial turbulent Mach number distributions $a_{n\ell m}$ in the Si/O-rich layer for models 25M, 22L, and 27\LA.
The key parameters of the Si/O-rich layer and the SSH decomposition are listed in Table \ref{tab:harm_param}.
Each decomposition component is calculated using Eq. \ref{aeq:components}.
The decomposition components in the range of $n = 0, 1, 2, ..., n_{\rm end}$, $\ell = 0, 1, 2, ..., \ell_{\rm end}$, and $m = 0, 1, 2, ..., \ell$ for each $\ell$ in models 25M, 22L, and 27\LA are listed in Tables \ref{tab:anlm_25M}, \ref{tab:anlm_22L}, and \ref{tab:anlm_27LA}, respectively. These data are also available as machine-readable tables.\footnote[7]{The structures of models 25M, 22L, and 27\LA at the last step of the 1D evolution calculations and a code to reconstruct the radial Mach number distribution from the SSH decomposition components are available at https://github.com/yosshiida/mndsb.
}
The negative $m$ components are calculated using the relations
\begin{eqnarray}
    {\rm Re}(a_{n\ell -m}) &=& (-1)^{m} {\rm Re}(a_{n\ell m}), \nonumber \\
    {\rm Im}(a_{n\ell -m}) &=& (-1)^{m+1} {\rm Im}(a_{n\ell m}),
\end{eqnarray}
where Re($z$) and Im($z$) are real and imaginary parts of a complex number $z$, respectively.
The radial turbulent Mach number at each point is reconstructed from the decomposed components with Eq. (\ref{aeq:decomposition}).

When one applies the decomposition components to the structure of a 1D model, one needs to fit the region of the convection layer to the corresponding region of the 1D model.
When the radii of the inner and outer boundaries are $r'_{\rm in}$ and $r'_{\rm out}$, the wavenumber of the radial component of the Helmholtz equation $k'_{n\ell}$ also changes to satisfy Dirichlet boundary conditions with
\begin{equation}
    g_{\ell}(k'_{n\ell}r'_{\rm in}) = g_{\ell}(k'_{n\ell}r'_{\rm out}) = 0.
\end{equation}
In order to keep the amplitude of the radial Mach number in the changed convection region, one can change the decomposition components $a'_{n\ell m}$ as
\begin{equation}
    a'_{n\ell m} = \left( \frac{{r^\prime}_{\rm out}^3 - {r^\prime}_{\rm in}^3}{r_{\rm out}^3 - r_{\rm in}^3} \right)^{\frac{1}{2}} a_{n\ell m}.
    \label{eq:mod-components}
\end{equation}
The radial Mach number is calculated using Eq. \ref{aeq:decomposition} in Appendix with the modified wavenumbers $k'_{n\ell}$ and the set of the decomposed components $a'_{n\ell m}$.
Note that the modified wavenumbers are not strictly scaled when the ratio $r_{\rm out}/r_{\rm in}$ changes.
However, the global feature of the radial Mach number distribution scarcely changes by the change of the $r_{\rm out}/r_{\rm in}$ ratio (see Fig. \ref{f10}).

\begin{deluxetable*}{lcc}[th]
\tablecaption{Summary Table of the Adopted Figures of the Spatiotemporal Evolution of the Angle-averaged Mass Fractions of O, Ne, and Si in Models 22L and 27\LA
\label{tab:figure_number}}
\tablewidth{0pt}
\tablehead{
\colhead{Component} & \colhead{Model 22L} & \colhead{Model 27\LA}
}
\startdata
O mass fraction & Figure \ref{fig:mforav22L} ($r_8$ = 0--15) & Figure \ref{fig:mforav27LA_Appendix} ($r_8$ = 0--70) \\
 & Figure \ref{fig:mforav22L_Appendix} ($r_8$ = 2--5) & Figure \ref{fig:mfrav27LA_SiOinner}, right panel ($r_8$ = 0--10) \\
Ne mass fraction & --- & Figure \ref{fig:mforav22L_Appendix}, top panel ($r_8$ = 0--70) \\
 & --- & Figure \ref{fig:mfner27LA}, bottom panel ($r_8$ = 0--10) \\
Si mass fraction & --- & Figure \ref{fig:mfrav27LA} ($r_8$ = 0--70) \\
 & --- & Figure \ref{fig:mfrav27LA_SiOinner}, left panel ($r_8$ = 0--10) \\
\enddata
\tablecomments{$r_8$ is the radius in units of $10^8$ cm.}
\end{deluxetable*}

\begin{figure*}
\vspace*{-1cm}
\includegraphics[width=18cm]{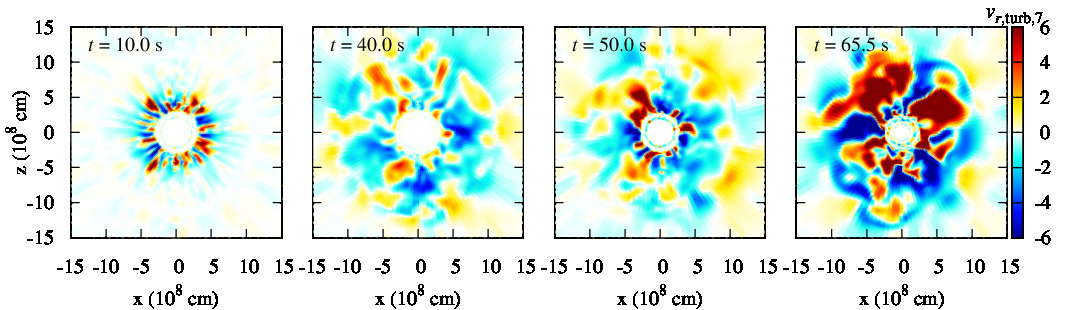}
\includegraphics[width=18cm]{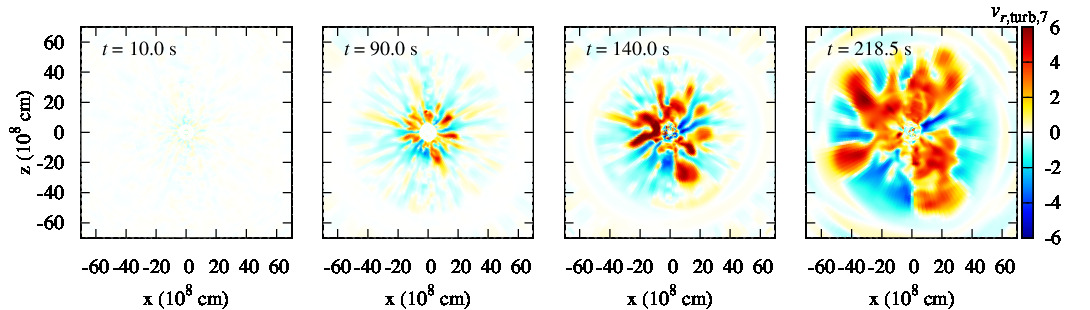}
\caption{2D slices on the \textit{x-z} plane showing radial turbulent velocity in the Si/O-rich layer.
Top panels indicate the radial turbulent velocity at $t = 10$, 40, 50, and 65.5 of model 22L from the left to the right.
Bottom panels indicate the same at $t = 10$, 90, 140, and 218.5 s of model 27\LA from left to right.
\label{fig:vrturb_xz}}
\end{figure*}

\begin{figure}[h]
\includegraphics[width=\columnwidth]{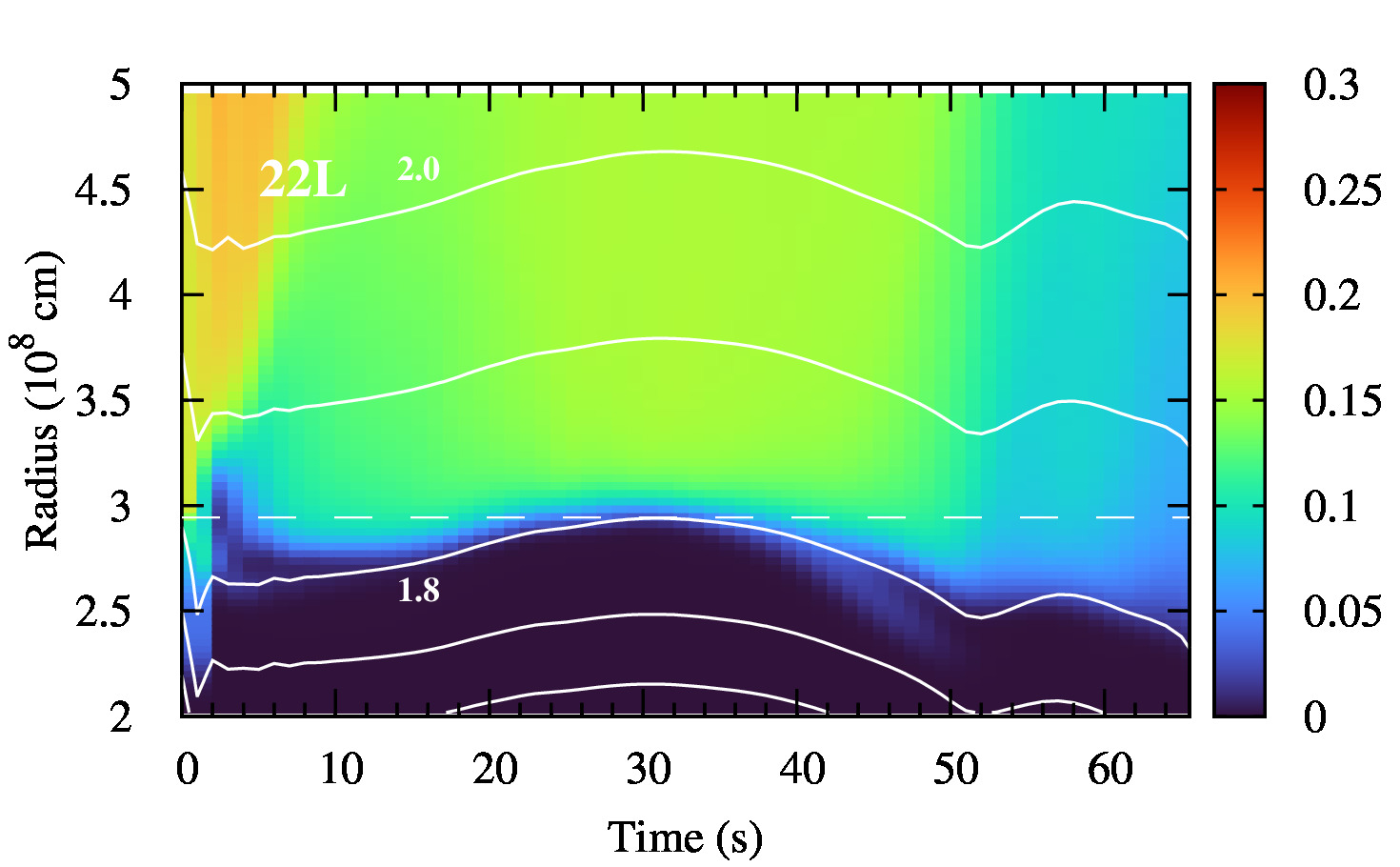}
\caption{Spatiotemporal evolution of the angle-averaged O mass fraction for model 22L.
The radial range is zoomed in close to the inner boundary ($2 \times 10^{8}$ cm $\le r \le 5 \times 10^{8}$ cm).
See Figure \ref{fig:mforav22L} for comparison.
The white dashed line indicates the initial radii of the inner boundary of the Si/O layer.
The white curves indicate the mass coordinates from 1.6 to 2.0 $M_\odot$ in intervals of 0.1 $M_\odot$.
\label{fig:mforav22L_Appendix}}
\end{figure}

\section{Additional figures for models 22L and 27\LA and a summary table of the spatiotemporal evolution of the mass fraction distributions.}

The final evolution of model 22L is similar to model 25M, which has been already presented in \citet{Yoshida19}.
In this appendix, we briefly summarize the evolution of model 22L for the sake of comparison with model 27\LA that we mainly focus on this work (e.g., Table \ref{tab:figure_number}).

Figure \ref{fig:vrturb_xz} provides 2D slices on the \textit{x-z} plane showing the radial turbulent velocity in the Si/O-rich layer for models 22L and 27\LA. 
For model 22L, turbulence starts with a small-scale velocity distribution in 10 s.
Then, the velocity fluctuation expands radially outward in 40 s.
The oxygen shell burning starts from $\sim$50 s and turbulence with a large-scale variation compared with that in $\sim$10 s develops. The top four panels of Figure 19 show that turbulence grows with time until the final simulation time ($\sim$65 s).
In the case of model 27\LA
 (bottom four panels), turbulence develops with a longer timescale compared with model 22L.
We see small-scale velocity fluctuations in 90 s.
The increase in the Ne mass fraction below $\sim$ 10$^{9}$ cm is caused by the downflows of the turbulence (see Figures \ref{fig:mfrav27LA_Ne} and \ref{fig:2Dmfne}).
In the phase III, turbulence with a large-scale variation develops  outward (140 s) and develops to the whole O/Si/Ne layer (218.5 s).

Model 22L has an Si/O convective layer activated by the oxygen shell burning similarly to model 25M.
Figure \ref{fig:mforav22L_Appendix} shows the time and radial profiles of the angle-averaged O mass fraction in the inner region of the Si/O layer ($2 \times 10^{8}$ cm $\le r \le 5 \times 10^{8}$ cm) of model 22L.
The Si/O layer initially expands by the oxygen shell burning and then contracts.
The contraction leads to the O shell burning again from $\sim$50 s.
This burning reduces the O mass fraction in this layer and causes a slight expansion before the collapse ($t \sim 65$ s).

\begin{figure*}[th]
\includegraphics[width=\textwidth]{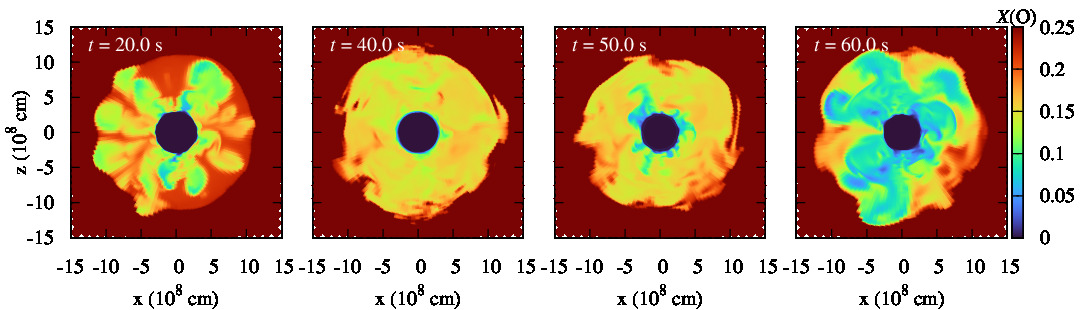}
\caption{Slices on the $xz$-plane showing the O mass 
fraction distribution of model 22L at 20, 40, 50, and 60 s from the left to the right.
The side length of the cubic box is $3.0 \times 10^{9}$ cm.
\label{fig:2Dmfo22L}}
\end{figure*}

\begin{figure*}[th]
\includegraphics[width=\textwidth]{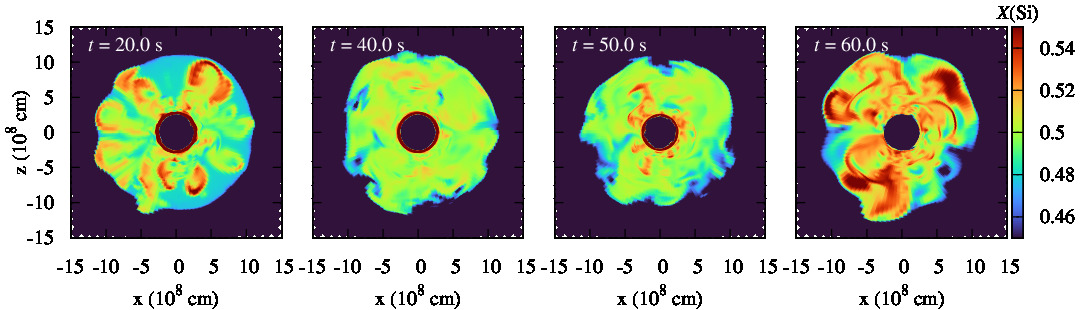}
\caption{Same as Figure \ref{fig:2Dmfo22L} but for the Si mass fraction.
\label{fig:2Dmfsi22L}}
\end{figure*}

\begin{figure}[th]
\vspace*{8mm}
\includegraphics[width=\columnwidth]{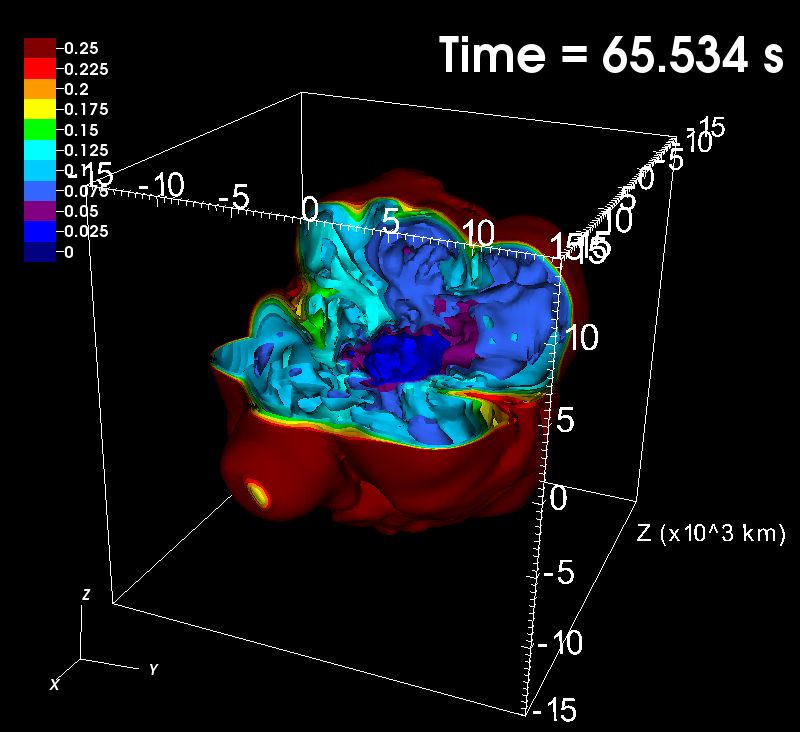}
\caption{The 3D contours the O mass 
fraction distribution of model 22L at 65.5 s. 
The side length of the cubic box is $3.0 \times 10^{9}$ cm.
\label{fig:3Dmfo22L}}
\end{figure}

\begin{figure}[h]
\vspace*{8mm}
\includegraphics[width=\columnwidth]{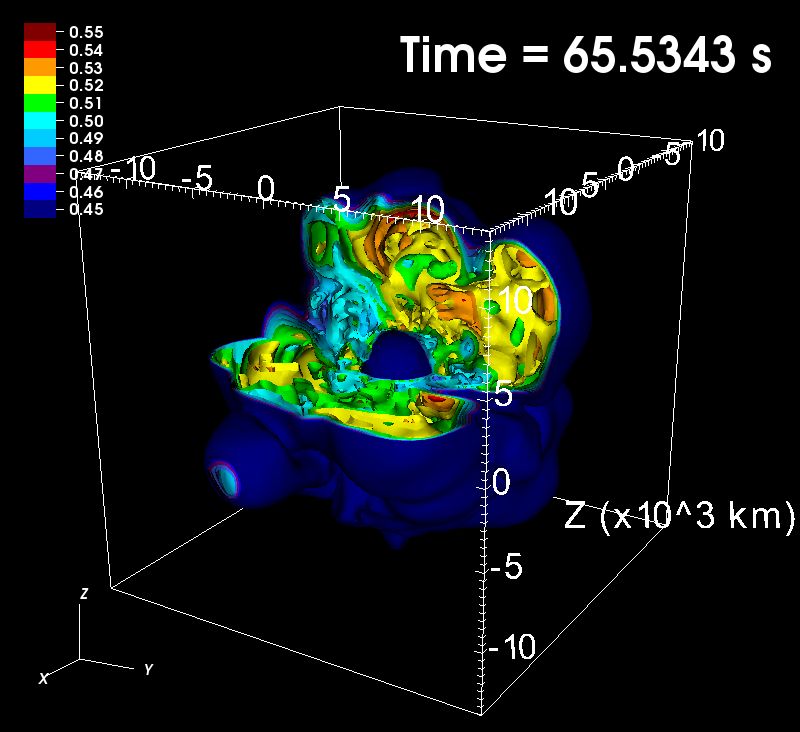}
\caption{Same as Figure \ref{fig:3Dmfo22L} but for the Si mass fraction.
\label{fig:3Dmfsi22L}}
\end{figure}

We show in Figures \ref{fig:2Dmfo22L} and \ref{fig:2Dmfsi22L} the 2D slices on the \textit{x-z} plane of the O and Si mass fraction distributions of model 22L, respectively, for the comparison to model 27\LA.
Figures \ref{fig:3Dmfo22L} and \ref{fig:3Dmfsi22L} correspond to Figure \ref{fig:3Dmfsi} for model 27\LA.
We see at 20 s upflows of O-poor and Si-rich plumes.
Then, the inhomogeneity of the O and Si mass fractions becomes smaller in $\sim$40 s.
From $\sim$50 s, the oxygen shell burning starts and the O-poor Si-rich materials go up again.
We see that the Si-rich materials extend to the Si/O layer, and this layer has not become homogeneous.

\begin{figure}
\includegraphics[width=\columnwidth]{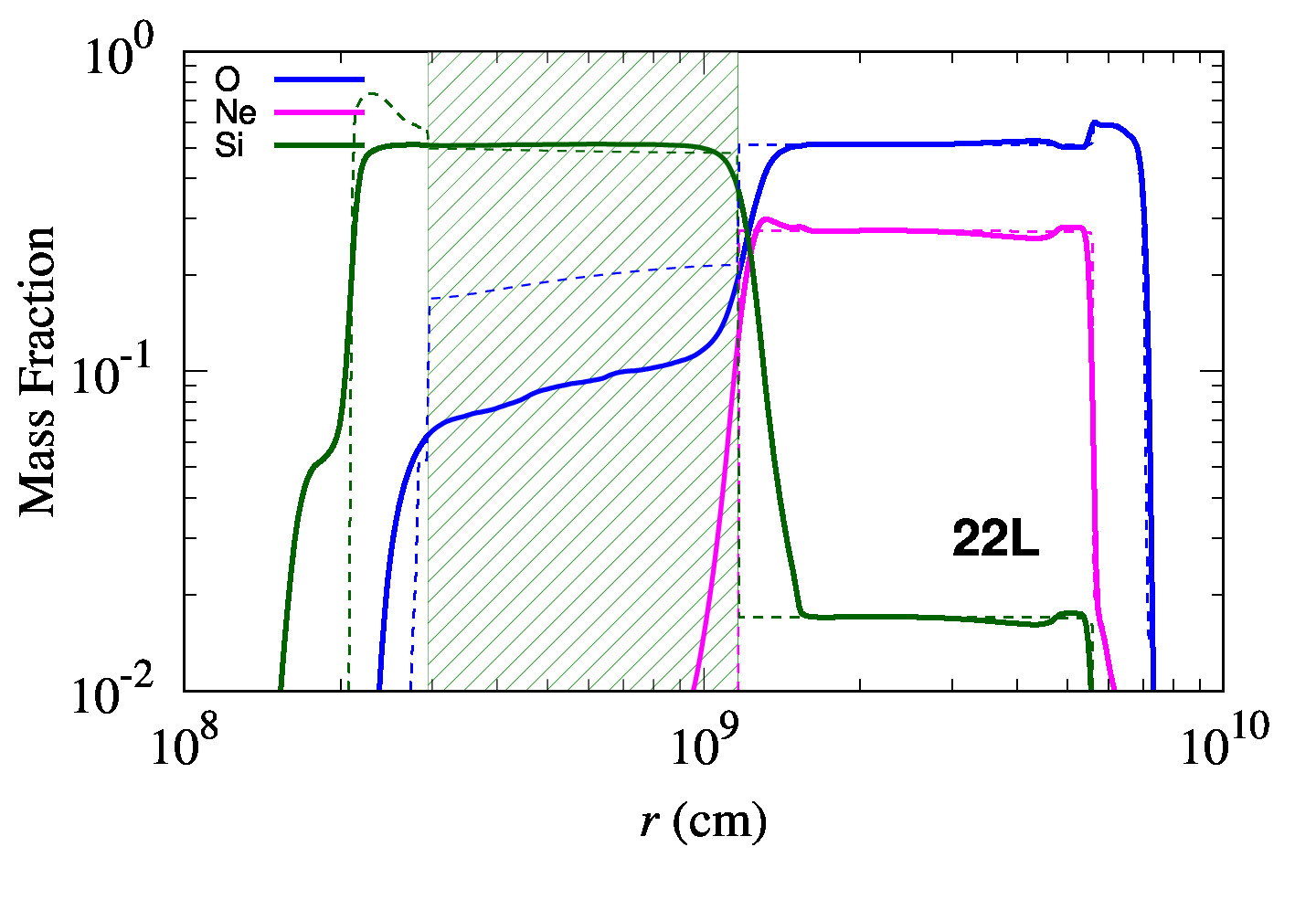}
\includegraphics[width=\columnwidth]{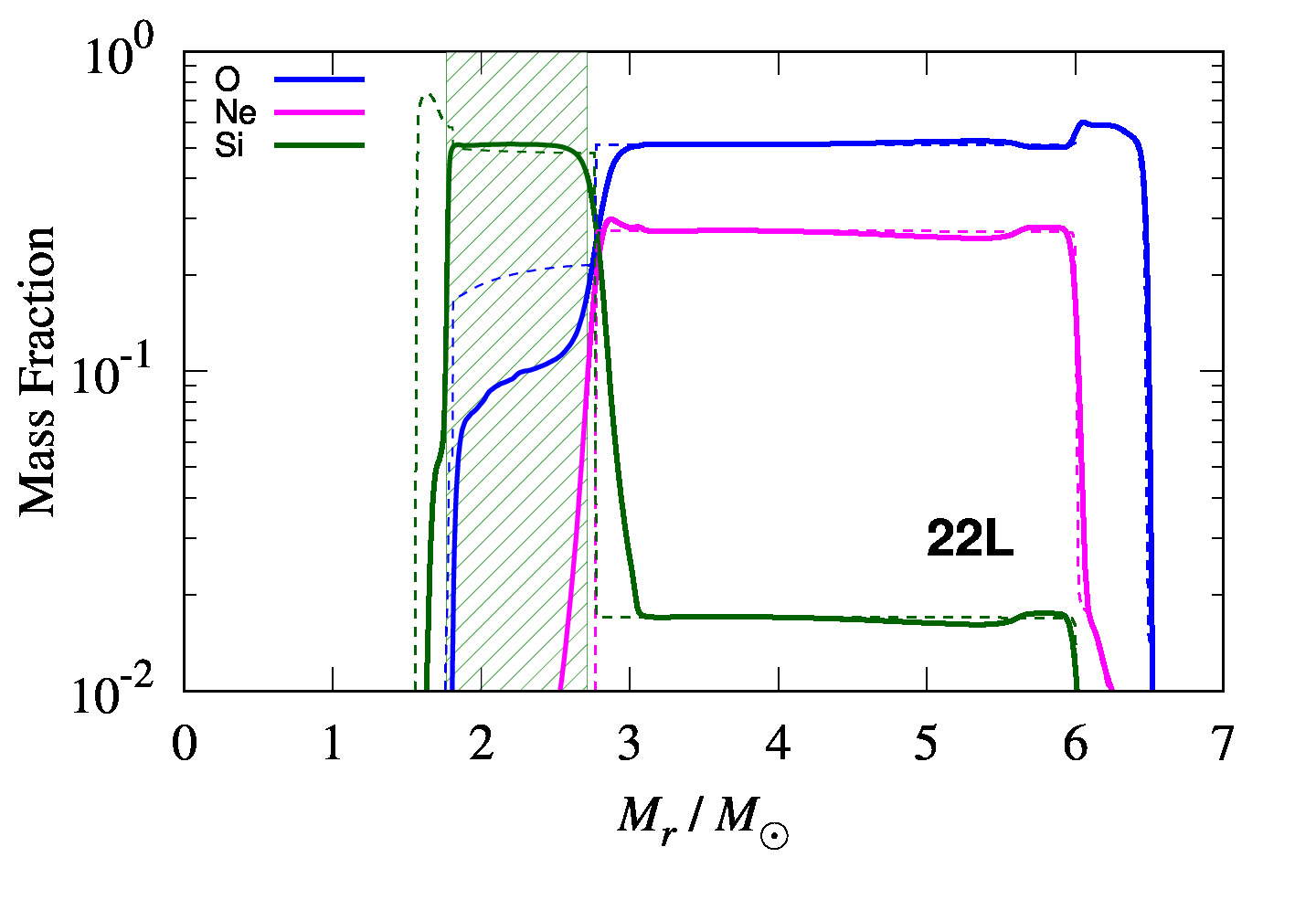}
\caption{Same as Figure \ref{fig:mfONeSi} but for model 22L.
\label{fig:mfONeSi_22L}}
\end{figure}

Figure \ref{fig:mfONeSi_22L} shows the angle-averaged mass fraction profiles of O, Ne, and Si at the last step of model 22L.
This figure corresponds to Figure \ref{fig:mfONeSi} for model 27\LA.
The O mass fraction in the Si/O layer at the last simulation time (solid blue line) is smaller than the initial mass fraction
 (dashed blue line).
The O mass fraction decreases inward to the bottom of the Si/O layer owing to the O burning. 
This is because the O burning proceeds in a shorter timescale than the mixing timescale that works to homogenize the composition distribution in the layer.

\begin{figure}[h]
\includegraphics[width=\columnwidth]{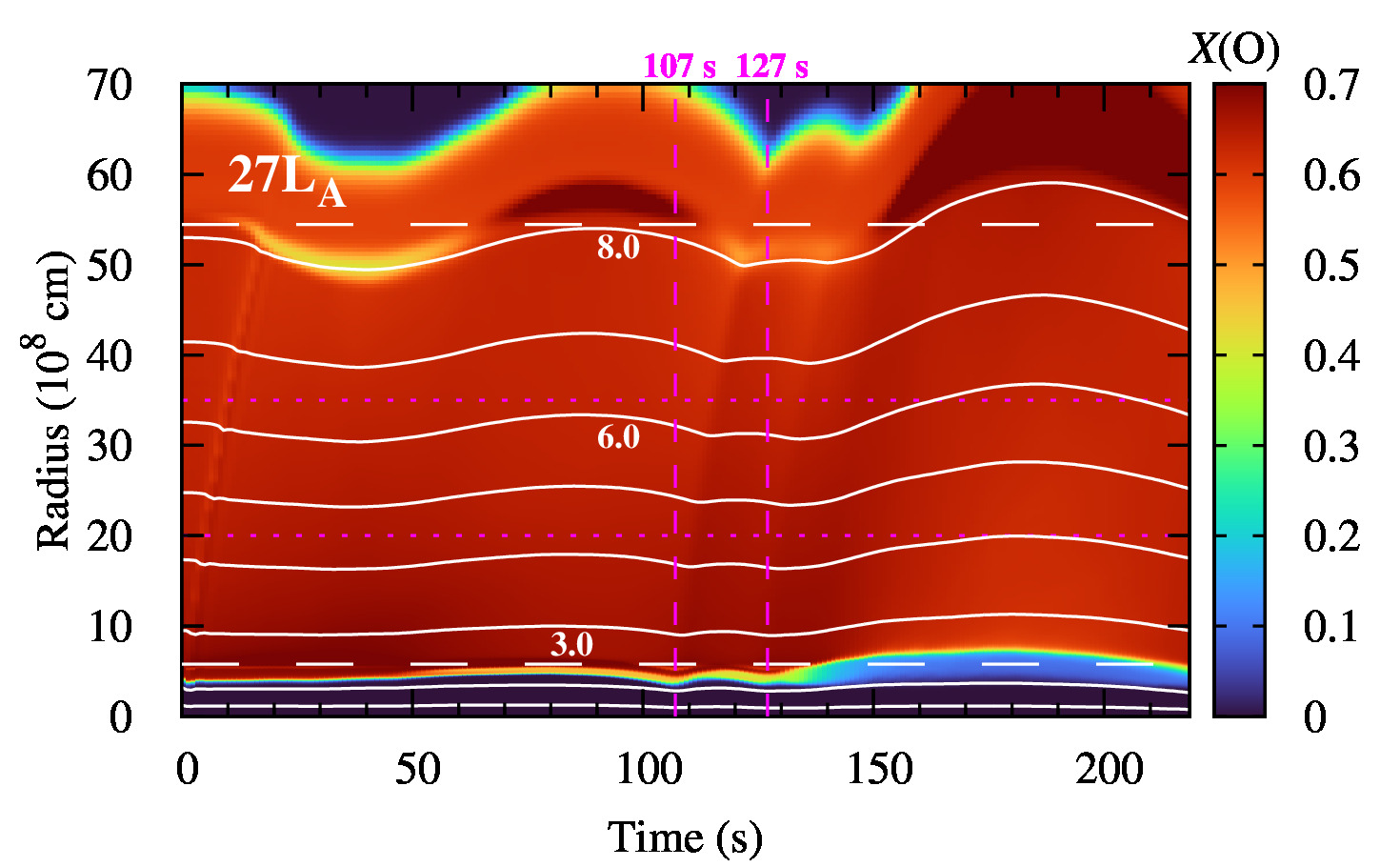}
\caption{Same as Figure \ref{fig:mfrav27LA} but for the O mass fraction.
\label{fig:mforav27LA_Appendix}}
\end{figure}

Finally, Figure \ref{fig:mforav27LA_Appendix} shows spatiotemporal evolution of the angle-averaged O mass fraction in the whole convective O/Si/Ne layer of model 27\LA, corresponding to the Si mass fraction shown in Figure \ref{fig:simass27LA}.
Although expansions and contractions of the O/Si/Ne layer are seen, we do not see remarkable changes in the O mass fraction distribution for this model.

\bibliographystyle{aasjournal}
\bibliography{ms.bbl}



\end{document}